\pdfoutput=1
\documentclass[a4paper,12pt]{article}
\usepackage{amsmath}
\usepackage{amssymb}
\usepackage{amsfonts}
\usepackage{graphicx}
\usepackage{xfrac}
\usepackage{bbold}

\usepackage[utf8]{inputenc}

\usepackage[square,numbers,comma,sort&compress]{natbib}
\usepackage[colorlinks=true,citecolor=blue,linkcolor=purple,urlcolor=purple]{hyperref}
\usepackage{geometry}
\usepackage[usenames,dvipsnames,svgnames,table]{xcolor}
\usepackage{booktabs}
\usepackage{slashed}

\def\varabstract{ }
\def\varkeywords{ }
\def\vararxivnumber{ }
\def\vartitle{ }
\def\varsubtitle{ }
\renewcommand{\title}[1]{\gdef\vartitle{#1}}

\renewcommand{\abstract}[1]{\gdef\varabstract{#1}}
\newcommand{\keywords}[1]{\gdef\varkeywords{#1}}

\newtoks\authtoks
\renewcommand{\author}[2][]{%
	\authtoks=\expandafter{\the\authtoks#2$^{#1}$\ }%
}
\newtoks\affiltoks
\newcommand{\affiliation}[2][]{%
    \affiltoks=\expandafter{\the\affiltoks{\item[$^{#1}$]#2}}%
}
\newtoks\emailtoks\newcounter{emailcounter}%
\newcommand{\emailAdd}[1]{%
\ifnum\theemailcounter>0\emailtoks=\expandafter{\the\emailtoks, \typeemail{#1}}%
\else\emailtoks=\expandafter{\typeemail{#1}}%
\fi
\stepcounter{emailcounter}}
\newcommand{\typeemail}[1]{\href{mailto:#1}{\tt #1}}

\renewcommand\maketitle{
	\newgeometry{margin=2cm}
	\pagestyle{empty}\setcounter{page}{0}
	{\huge\flushleft\sffamily\bfseries\vartitle\\\Large\varsubtitle\par}
\vskip6ex
{\large\bfseries\raggedright\sffamily\the\authtoks\par}
\vskip2ex
\begin{list}{}{%
\setlength{\leftmargin}{0.28cm}%
\setlength{\labelsep}{0pt}%
\setlength{\itemsep}{-3pt}%
\setlength{\topsep}{-\parskip}}
\itshape\small%
\the\affiltoks
\end{list}
\vskip2ex
\noindent\hspace{0.28cm}\begin{minipage}[l]{.9\textwidth}
\begin{flushleft}
\textit{E-mail:} \the\emailtoks
\end{flushleft}
\end{minipage}
\vskip5ex
\noindent{\renewcommand\baselinestretch{.9}\textsc{Abstract:}}\ \varabstract
\vskip5ex
\if!\varkeywords!\else\noindent{\textsc{Keywords:}}\ \varkeywords \vskip2ex\fi
\if!\vararxivnumber!\else\noindent{\textsc{ArXiv ePrint: 1903.01242}}
\href{http://arxiv.org/abs/\vararxivnumber}{\vararxivnumber}\vskip2ex\fi

\newpage
\restoregeometry
\pagestyle{plain}

\setcounter{footnote}{0}
}

\usepackage[square,numbers,comma,sort&compress]{natbib}

\definecolor{MS}{rgb}{0,0,1}

\usepackage{ifthen}

	\newcommand{\barlimc}[7]{
  \pgfmathparse{\mypos+0.3}
  \edef\mypos{\pgfmathresult}
		\node[left,scale=0.6] at (0,\mypos) {#1};
		\pgfmathparse{#3 > 5 ? 1 : 0}
		\ifthenelse{\pgfmathresult=1}{
			\fill[#2] ($(0,\mypos)+(0,-0.1)$) rectangle +(5,0.2);
			\fill[white] ($(0,\mypos)+(3.5,-0.1)$) rectangle +(0.3,0.2);
			\draw[decoration={zigzag},decorate,#2,very thick] (3.4,\mypos) to +(0.5,0);
			\node[left,scale=0.6] at (5,\mypos) {#3};
			}{
			\fill[#2] ($(0,\mypos)+(0,-0.1)$) rectangle +(#3,0.2);
			\node[left,scale=0.6] at (#3,\mypos) {#3};
		}
		\fill[#4] ($(0,\mypos)+(0,-0.1)$) rectangle +(#5,0.2);
		\node[left,scale=0.6] at (#5,\mypos) {#5};
		\fill[#6] ($(0,\mypos)+(0,-0.1)$) rectangle +(#7,0.2);
		\pgfmathparse{#7 <0.3 ? 1 : 0}
		\ifthenelse{\pgfmathresult=1}{
			\node[right,scale=0.6] at (0,\mypos) {#7};
		}{
		\node[left,scale=0.6] at (#7,\mypos) {#7};
	}
}

\title{\hspace{\fill}\mbox{\footnotesize\rm NCTS-PH/1902}\bigskip\bigskip\\
Charged-lepton-flavor violation in \boldmath$|\Delta S|=1$ hyperon decays}

\author[1,2,3]{Xiao-Gang He,}\emailAdd{hexg@phys.ntu.edu.tw}
\author[1,3]{ Jusak Tandean}\emailAdd{jtandean@phys.ntu.edu.tw}
\author[4]{ and German Valencia}\emailAdd{german.valencia@monash.edu}

\affiliation[1]{Department of Physics, National Taiwan University,\\
No.\,\,1, Sec.\,\,4, Roosevelt Rd., Taipei 106, Taiwan}
\affiliation[2]{Tsung-Dao Lee Institute $\&$ Department of Physics and Astronomy, SKLPPC,
Shanghai Jiao Tong University, 800 Dongchuan Rd., Minhang, Shanghai 200240, China}
\affiliation[3]{Physics Division, National Center for Theoretical Sciences,\\
No.\,\,101, Sec.\,\,2, Kuang Fu Rd., Hsinchu 300, Taiwan}
\affiliation[4]{School of Physics and Astronomy, Monash University, Melbourne VIC-3800, Australia}

\abstract{Studies of lepton-flavor violation in strangeness-changing ($|\Delta S|=1$) transitions have a~long tradition in the kaon sector where they provide some of the strongest limits on physics beyond the standard model. Recent hints of violation of lepton-flavor universality in $B$-meson decays have revived interest in lepton-flavor violation as the two phenomena appear simultaneously in many extensions of the standard model. At the same time, the LHCb experiment has produced new results for the hyperon process $\Sigma^+\to p\mu^+\mu^-$ and may be in a position to study other rare hyperon decay modes. With this in mind, we investigate $|\Delta S|=1$ hyperon decays into different-flavor lepton pairs $e^\pm\mu^\mp$ in a model-independent manner and contrast the coverage of parameter space that can be achieved with what is known from kaon modes. We include a comparison with selected two-body leptonic decays of charged mesons, with $K\to \pi\nu\bar\nu$ modes, with $\mu\to e$ conversion, and with lepton-flavor violating decays of other neutral mesons,  all of which constrain the same parameter space in a complementary~way.
}


\begin{document}
\baselineskip=17pt \parskip=5pt

\maketitle

{\hypersetup{linkcolor=black}
  \tableofcontents}

\newpage

Charged-lepton-flavor violation (LFV) occurs within the standard model (SM) when neutrino masses are included, but since these masses are extremely small, the resulting LFV is strongly suppressed. For this reason, processes manifesting LFV provide an ideal window to new physics~(NP), and hence quests for them are of tremendous importance.
Many extensions of the SM do not preserve lepton-flavor number, and the corresponding parameters have been tightly restricted by the negative outcomes of the various searches conducted so far in the decays of kaons, $B$ mesons, and charged leptons, among others~\cite{Tanabashi:2018oca,Buchmuller:1985jz,Littenberg:1993qv,Raidal:2008jk,Cai:2015poa,Lindner:2016bgg,Carpentier:2010ue,Lee:2015qra,Crivellin:2016vjc,Hazard:2017udp,Davidson:2018rqt}. 
The most common examples of NP exhibiting LFV include leptoquarks~\cite{Davies:1990sc,Davidson:1993qk,Valencia:1994cj,Cheung:2015yga,Baek:2015mea,Dorsner:2016wpm,Hiller:2016kry,Fayyazuddin:2018zww,Sahoo:2018ffv,deMedeirosVarzielas:2018bcy,Kim:2018oih}, heavy neutrinos~\cite{Cheng:1976uq,Bilenky:1977du,Bjorken:1977br,Cvetic:2002jy,He:2004wr,Lee:2014rba,He:2015rqa,Lami:2016mjf,Rocha-Moran:2016enp,Zhang:2018nmy}, gauged U(1) extensions of the SM with their associated $Z^\prime$ gauge bosons~\cite{Langacker:2000ju,Chiang:2011cv,Crivellin:2015era,Chiang:2016qov,Kim:2016bdu,Chiang:2017hlj,Aranda:2018zis,Mu:2018weh,Ko:2019tts}, and multi-Higgs models~\cite{Bjorken:1977vt,Botella:2015hoa,Baek:2015fma,Benbrik:2015evd,Omura:2015xcg,Sher:2016rhh,Wang:2016rvz}.
Interestingly, some of these NP possibilities can give rise to lepton-flavor-universality violations of the type hinted at by recent $B$-physics measurements of the quantities $R_{K^{(\star)}}$ and $R_{D^{(\star)}}$ \cite{Tanabashi:2018oca,Aaij:2014ora}.

Tests of LFV in strangeness-changing ($|\Delta S|=1$) quark transitions have a long tradition in kaon physics where the experimental branching-fraction limit ${\cal B}(K_L\to e^\pm\mu^\mp)<4.7\times 10^{-12}$ \cite{Ambrose:1998us} can be interpreted as probing energy scales above 100~TeV \cite{Tanabashi:2018oca}.
Only slightly less impressive are the constraints that have been obtained from the $K\to\pi e^\pm\mu^\mp$ modes.
There are no corresponding limits from the light hyperon sector as far as we know, but the recent measurement by the LHCb Collaboration of \,${\cal B}(\Sigma^+\to p \mu^+\mu^-)=\big(2.2^{+1.8}_{-1.3}\big)\times 10^{-8}$ \cite{Aaij:2017ddf} suggests that new limits from this sector could become available soon.

Our purpose in this paper is to explore LFV in \,$|\Delta S|=1$\, hyperon decays in a model-independent way and to compare the coverage of NP parameter space they offer to that already available from kaon studies. Our work is partly motivated by the ongoing efforts by LHCb to investigate hyperon and kaon processes \cite{Borsato:2018tcz,Junior:2018odx,Bediaga:2018lhg}.

The organization of this paper is as follows. In section \ref{Leff} we consider the most general effective Lagrangian involving quark-lepton operators of  dimension six which are invariant under the SM gauge group and can induce \,$|\Delta S|=1$\, transitions with LFV among the lightest hadrons.  We then briefly discuss a possible ultraviolet completion of this effective Lagrangian in terms of leptoquarks. In section~\ref{drates}, we first obtain the baryonic matrix elements pertaining to our hyperon decays of interest and subsequently derive their decay rates. We also deal with their kaon counterparts as well as other processes without hyperons that are affected by the same operators as a consequence of gauge invariance, such as \,$K\to\pi\nu\bar\nu$\, and \,$\mu\to e$\, conversion in nuclei. In section~\ref{numeric}, we present our numerical analysis and illustrate how the different processes are complementary in probing the NP of concern.
We summarize and draw our conclusions in section \ref{conclusions}.
Some technical details are relegated to appendices.

\section{Effective Lagrangian\label{Leff}}

\subsection{Model-independent approach\label{general}}

We begin from the most general effective Lagrangian that can be built out of SM fields, including gauge fields and a light Higgs,\footnote{I.e., the linear realization of electroweak symmetry breaking.\medskip} and respects the gauge symmetries of the SM, as has been described before in the literature \cite{Buchmuller:1985jz,Grzadkowski:2010es}.
The operators ${\cal Q}_k^{}$ that can contribute to \,$|\Delta S|=1$\, transitions with LFV between down-type light fermions first occur at dimension six (dim-6).
There are several such operators \cite{Buchmuller:1985jz,Grzadkowski:2010es}, and the Lagrangian containing them has the form
\begin{equation} \label{Lnpo}
{\cal L}_{\textsc{np}}^{} \,=\, \frac{1}{\Lambda_{\textsc{np}}^2} \Bigg[
\raisebox{1pt}{\footnotesize$\displaystyle\sum_{\mbox{\scriptsize$k=1$}}^{\mbox{\scriptsize5}}$}\, {\cal C}_k^{ijxy} {\cal Q}_k^{ijxy}
+ \big( {\cal C}_6^{ijxy} {\cal Q}_6^{ijxy} + {\rm H.c.} \big) \Bigg] \,,
\end{equation}
where $\Lambda_{\textsc{np}}$ denotes a heavy mass scale characterizing the underlying NP interactions, ${\cal C}_{1,...,6}^{ijxy}$ are dimensionless and generally complex coefficients, \,$i,j,x,y=1,2,3$\, stand for family indices, summation over them being implicit,
\begin{align} \label{Qset}
{\cal Q}_1^{ijxy} & \,=\, \overline{q_i^{}} \gamma^\eta q_j^{}\, \overline{l_x^{}} \gamma_\eta^{} l_y^{} \,, &
{\cal Q}_2^{ijxy} & \,=\, \overline{q_i^{}} \gamma^\eta \tau_{\texttt I}^{} q_j^{}\, \overline{l_x^{}} \gamma_\eta^{} \tau_{\texttt I}^{} l_y^{} \,, &
{\cal Q}_3^{ijxy} & \,=\, \overline{d_i^{}}\gamma^\eta d_j^{}\, \overline{e_x^{}}\gamma_\eta^{} e_y^{} \,,
\nonumber \\
{\cal Q}_4^{ijxy} & \,=\, \overline{d_i^{}} \gamma^\eta d_j^{}\, \overline{l_x^{}} \gamma_\eta^{} l_y^{} \,, &
{\cal Q}_5^{ijxy} & \,=\, \overline{q_i^{}} \gamma^\eta q_j^{}\, \overline{e_x^{}} \gamma_\eta^{} e_y^{} \,, &
{\cal Q}_6^{ijxy} & \,=\, \overline{l_i^{}} e_j^{}\, \overline{d_x^{}} q_y^{} \,, &
\end{align}
with $q_i$ and $l_i^{}$ ($d_i^{}$ and $e_i^{}$) representing the left-handed doublets (right-handed singlets) of quarks and leptons, respectively, $\tau_{\texttt I=1,2,3}^{}$ denoting Pauli matrices, and $\texttt I$ being summed over.\footnote{Hence \,${\cal Q}_1^{}=Q_{lq}^{(1)}$, \,${\cal Q}_2^{}=Q_{lq}^{(2)}$, \,${\cal Q}_3^{}=Q_{ed}^{}$, \,${\cal Q}_4^{}=Q_{ld}^{}$, \,${\cal Q}_5^{}=Q_{eq}^{}$,\, and \,${\cal Q}_6^{}=Q_{leqd}^{}$\, in the notation of~\cite{Grzadkowski:2010es}.\medskip}
Accordingly, \,${\cal Q}_k^{ijxy\dagger}={\cal Q}_k^{jiyx}$,\, and also \,${\cal C}_k^{ijxy*}={\cal C}_k^{jiyx}$\, due to the Hermiticity of ${\cal L}_{\textsc{np}}$, for \,$k=1,...,5$.\,
We note that there are no dim-6 SM-gauge-invariant operators comprising tensor bilinears that directly participate in down-type quark-lepton transitions, as previously observed \cite{Grzadkowski:2010es,Alonso:2014csa,Bardhan:2017xcc}.\footnote{Since \,$\overline{\texttt f_1}\sigma_{\kappa\eta}^{}(1-\gamma_5^{})\texttt f_2\,\overline{\texttt f_3}\sigma^{\kappa\eta}(1+\gamma_5^{})\texttt f_4=0$\, for any fermion fields $\texttt f_{1,2,3,4}$, the only dim-6 SM-gauge-invariant tensor-bilinear product is $\varepsilon_{ac}^{}\big(\bar l_i^a\sigma_{\kappa\eta}^{}e_j\big)\big(\bar q_x^c\sigma^{\kappa\eta}u_y\big)$, where the weak-isospin indices \,$a,c=1,2$\, are summed over, \,$\varepsilon_{12}^{}=-\varepsilon_{21}^{}=1$,\, $\varepsilon_{11,22}^{}=0$,\, and $u_y$ is a right-handed up-type quark field \cite{Grzadkowski:2010es}.\medskip}
Moreover, the absence of tree-level flavor-changing neutral currents in the SM implies that dim-6 operators made up of a quark or lepton bilinear in combination with gauge and Higgs fields also do not contribute to ${\cal L}_{\textsc{np}}$.

For convenience, we can choose to work in the mass basis of the down-type fermions, where
\begin{align} \label{ql}
q_i^{} & \,=\, P_L^{}\! \left(\!\begin{array}{c} \raisebox{1pt}{\footnotesize$\sum$}_j^{}
\big({\cal V}_{\textsc{ckm}}^\dagger\big)_{ij} \textsl{\texttt U}_j^{} \\ \textsl{\texttt D}_i^{} \end{array}\!\right) , &
l_i^{} & \,=\, P_L^{}\! \left(\!\begin{array}{c} \raisebox{1pt}{\footnotesize$\sum$}_j^{}
({\cal U}_{\textsc{pmns}})_{ij}^{}\nu_j^{} \\ \textsl{\texttt E}_i^{} \end{array}\!\right) , &
e_i^{} & \,=\, P_R^{} \textsl{\texttt E}_i^{} \,, & d_i^{} & \,=\, P_R^{} \textsl{\texttt D}_i^{} \,,
\end{align}
with \,${\cal V}_{\textsc{ckm}}$\, $({\cal U}_{\textsc{pmns}})$ being the Cabibbo-Kobayashi-Maskawa quark (Pontecorvo-Maki-Nakagawa-Sakata neutrino) mixing matrix, \,$P_{L,R}=(1\mp\gamma_5^{})/2$,\, and $\textsl{\texttt U}_{1,2,3}=u,c,t$, \,$\textsl{\texttt D}_{1,2,3}=d,s,b$, \,$\nu_{1,2,3}$,\, and \,$\textsl{\texttt E}_{1,2,3}=e,\mu,\tau$\, referring to the mass eigenstates.
We can then express the part of ${\cal L}_{\textsc{np}}$ containing operators $Q_k^{e\mu}$ and $Q_k^{\mu e}$ which contribute to \,$s\to d$\, transitions and do not conserve electron and muon flavors as
\begin{equation} \label{Lnp}
{\cal L}_{\textsc{np}} \,\supset\, \frac{1}{\Lambda_{\textsc{np}}^2}\,
\raisebox{1pt}{\footnotesize$\displaystyle\sum_{\mbox{\scriptsize$k=1$}}^{\mbox{\scriptsize6,6$\prime$}}$}
\big(c_k^{e\mu} Q_k^{e\mu}+c_k^{\mu e} Q_k^{\mu e}\big) \,,
\end{equation}
where \,$Q_k^{e\mu(\mu e)}={\cal Q}_k^{1212(1221)}$\, and \,$c_k^{e\mu(\mu e)}={\cal C}_k^{1212(1221)}$\, for \,$k=1,...,5$,\, while \,$Q_6^{e\mu}={\cal Q}_6^{1212}=\overline{l_1^{}}e_2^{}\,\overline{d_1^{}}q_2^{}$, \,$Q_6^{\mu e}={\cal Q}_6^{2112}$,\, $c_6^{e\mu(\mu e)}={\cal C}_6^{1212(2112)}$,\, $Q_{6\prime}^{e\mu}=\big({\cal Q}_6^{2121}\big){}^\dagger=\overline{q_1^{}}d_2^{}\,\overline{e_1^{}}l_2^{}$, \,$Q_{6\prime}^{\mu e}=\big({\cal Q}_6^{1221}\big){}^\dagger$, and~\,$c_{6\prime}^{e\mu(\mu e)}={\cal C}_6^{2121(1221)*}$.\,
The Hermitian conjugates of these terms are responsible for the corresponding \,$d\to s$\, transitions.
Given that the tau lepton is too heavy to appear in the final states of light hyperon decays, we do not discuss operators with the tau field.

For our study of hyperon processes and comparison with their kaon counterparts, it is convenient to rewrite eq.\,(\ref{Lnp}) explicitly separating parity-even and parity-odd quark couplings as
\begin{align} \label{s->dem}
{\cal L}_{\textsc{np}}^{} \,\supset\, \frac{-1}{\Lambda_{\textsc{np}}^2}~
\raisebox{4pt}{\footnotesize$\displaystyle\sum_{\mbox{\scriptsize$\ell,\ell'$}}$}\, & \Big[ \overline{d} \gamma^\kappa s~ \overline{\ell} \gamma_\kappa^{} \big( \texttt V_{\ell\ell'}^{} + \gamma_5^{}
\texttt A_{\ell\ell'}^{}\big) \ell'
+ \overline{d} \gamma^\kappa\gamma_5^{} s\, \overline{\ell} \gamma_\kappa^{} \big( \tilde{\textsc v}_{\ell\ell'}^{} + \gamma_5^{} \tilde{\textsc a}_{\ell\ell'}^{} \big) \ell'
\nonumber \\ & +\,
\overline{d} s\, \overline{\ell} \big( \texttt S_{\ell\ell'}^{} + \gamma_5^{}
\texttt P_{\ell\ell'}^{} \big) \ell'^{}
+ \overline{d} \gamma_5^{}s\, \overline{\ell} \big( \tilde{\textsc s}_{\ell\ell'}^{} + \gamma_5^{} \tilde{\textsc p}_{\ell\ell'}^{} \big)
\ell'^{} \Big] \,+\, {\rm H.c.} \,, &
\end{align}
where \,$\ell^{(\prime)}=e,\mu$\, but \,$\ell\neq\ell'$\, and $\texttt V_{\ell\ell'}$, $\texttt A_{\ell\ell'}$, $\texttt S_{\ell\ell'}$, $\texttt P_{\ell\ell'}$, $\tilde{\textsc v}_{\ell\ell'}$, $\tilde{\textsc a}_{\ell\ell'}$, $\tilde{\textsc s}_{\ell\ell'}$, and $\tilde{\textsc p}_{\ell\ell'}$ are dimensionless constants which can be complex.
As will be seen later on, in the rates of the hyperon and kaon decays of interest $\texttt V_{\ell\ell'}$, $\texttt A_{\ell\ell'}$, $\texttt S_{\ell\ell'}$, and $\texttt P_{\ell\ell'}$, which accompany the parity-even quark bilinears in~eq.\,(\ref{s->dem}), have no interference with $\tilde{\textsc v}_{\ell\ell'}$, $\tilde{\textsc a}_{\ell\ell'}$, $\tilde{\textsc s}_{\ell\ell'}$, and~$\tilde{\textsc p}_{\ell\ell'}$, which are associated with the parity-odd quark bilinears.
These couplings are related to the coefficients defined in eq.\,(\ref{Lnp}) by
\begin{align} \label{va}
4\, \texttt V_{\ell\ell'}^{} & \,=\, -c_1^{\ell\ell'} - c_2^{\ell\ell'} - c_3^{\ell\ell'} - c_4^{\ell\ell'} - c_5^{\ell\ell'} \,, &
4\, \texttt A_{\ell\ell'}^{} & \,=\,  c_1^{\ell\ell'} + c_2^{\ell\ell'} - c_3^{\ell\ell'} + c_4^{\ell\ell'} - c_5^{\ell\ell'} \,,
\nonumber \\
4\, \tilde{\textsc v}_{\ell\ell'}^{} & \,=\,  c_1^{\ell\ell'} + c_2^{\ell\ell'} - c_3^{\ell\ell'} - c_4^{\ell\ell'} + c_5^{\ell\ell'} \,, &
4\, \tilde{\textsc a}_{\ell\ell'}^{} & \,=\, -c_1^{\ell\ell'} - c_2^{\ell\ell'} - c_3^{\ell\ell'} + c_4^{\ell\ell'} + c_5^{\ell\ell'} \,, &
\\ \label{sp} \vphantom{|^{\int_|^|}}
4\, \texttt S_{\ell\ell'}^{} & \,=\, -c_6^{\ell\ell'} - c_{6\prime}^{\ell\ell'} \,=\, -4\, \tilde{\textsc p}_{\ell\ell'}^{} \,, &
4\, \texttt P_{\ell\ell'}^{} & \,=\, -c_6^{\ell\ell'} + c_{6\prime}^{\ell\ell'} \,=\, -4\, \tilde{\textsc s}_{\ell\ell'}^{} \,.
\end{align}
For $c_k^{\ell\ell'}$ being free parameters, $\texttt V_{\ell\ell'}$, $\texttt A_{\ell\ell'}$,
$\tilde{\textsc v}_{\ell\ell'}$, and $\tilde{\textsc a}_{\ell\ell'}$ are therefore linearly independent, whereas only two of their (pseudo)scalar partners are, which may be taken to be
$\texttt S_{\ell\ell'}$ and $\texttt P_{\ell\ell'}$.
We notice from eq.\,(\ref{sp}) that \,$c_6^{\ell\ell'}\pm c_{6\prime}^{\ell\ell'}$\, can each accompany both parity-even and parity-odd quark bilinears, which can of course also be understood
from the  explicit expressions for the relevant parts of $Q_{6,6\prime}^{\ell\ell'}$ in, say,
the \,$\ell\ell'=e\mu$\, case: \,$Q_6^{e\mu}\supset\overline{d}P_L^{}s\,\overline{e}P_R^{}\mu$\,
and \,$Q_{6\prime}^{e\mu}\supset\overline{d}P_R^{}s\,\overline eP_L^{}\mu$.\,

To see what other processes can receive contributions from the operators $Q_k^{e\mu,\mu e}$ in eq.\,(\ref{Lnp}), as well as their Hermitian conjugates, in appendix~\ref{feynrules} we summarize the Feynman rules that follow from them.
We then see that the decays listed below can also constrain these NP couplings.
Changes in lepton-flavor number can take place in all of these modes, and some of them involve one or two neutrinos.
\begin{itemize}
\item
$K^+ \to \pi^+ \nu \bar \nu$ and $K_L \to \pi^0 \nu \bar \nu$.
In this case, the NP contributions from $Q_k^{\ell\ell'}$ have no interference with the SM ones, due to their differing lepton-flavor combinations, but cause the decay rates to rise above the SM expectations, as the neutrinos are not detected.
\item
$\pi^\pm \to \ell^\pm \nu$,\, $K^\pm \to \ell^\pm \nu$,\, and \,$D_{(s)}^\pm \to \ell^\pm \nu$.\,
Since these are helicity suppressed in the~SM, the impact of physics beyond it will be most important on the electron modes, \,$\ell=e$.\,
Again, the NP represented by $Q_k^{\ell\ell'}$ does not interfere with the SM in these processes because it produces the `wrong' neutrino flavor. Since the neutrino flavor is not observed experimentally, these new contributions also increase the rates over their SM value.
\item
$\mu\to e$ conversion in nuclei and flavor-violating \,$\pi^0,\eta,D^0\to \ell^+\ell^{\prime -}$.
These serve as additional null tests of the SM, and there are ongoing searches for them.

\end{itemize}

It is worth remarking that among the operators ${\cal Q}_{1,...,6}^{ijxy}$ in eq.\,(\ref{Qset}) there are those not pertinent to $dse\mu$ interactions which can generally also influence some of the others listed in table\,\,\ref{tabFR}.
For instance, ${\cal Q}_1^{1112}$ in our mass basis, specified by eq.\,(\ref{ql}), contributes to $(\bar u u,\bar u c)\big(\bar e\mu,\bar\nu_e \nu_\mu\big)$ couplings.\footnote{Furthermore, there are operators \cite{Buchmuller:1985jz,Grzadkowski:2010es} not listed in eq.\,(\ref{Qset}) which contribute to these same couplings, such as \,$Q_{lu}^{ij12}=\overline{u_i^{}}\gamma^\eta u_j^{}\, \overline{l_1^{}}\gamma_\eta^{}l_2^{}$.}
In addressing the constraints from the preceding extra processes, we will ignore these other operators. This may be regarded as an implicit model assumption in our analysis.

\subsection{Leptoquark model}

To illustrate how ${\cal L}_{\textsc{np}}$ may be generated by renormalizable NP interactions,
we look at the leptoquark (LQ) scenario.
Amongst those that have been explored in the literature~\cite{Davies:1990sc,Davidson:1993qk,Valencia:1994cj,Cheung:2015yga,Baek:2015mea,Dorsner:2016wpm,Hiller:2016kry,Fayyazuddin:2018zww,Sahoo:2018ffv,deMedeirosVarzielas:2018bcy,Kim:2018oih}, with couplings to SM fermions which conserve baryon and lepton numbers and respect SM gauge symmetries, the LQs (with their ${\rm SU}(3)_{\textsc c}\times{\rm SU}(2)_L\times{\rm U}(1)_Y$ assignments) that can bring about ${\cal L}_{\textsc{np}}$ are $S_1\, \big(\bar 3,1,4/3\big)$, $S_2\, (3,2,7/6)$, $\tilde S_2\, (3,2,1/6)$, and $S_3\, \big(\bar 3,3,1/3\big)$, which are spinless, and $V_1\, (3,1,2/3)$, $V_2\, \big(\bar 3,2,5/6\big)$, and $V_3\, (3,3,2/3)$, which have spin 1.
The SU(2)$_L$ doublets (triplets) $S_2$, $\tilde S_2$, and $V_2$ ($S_3$ and $V_3$) each have two (three) components having different electric charges.
We can write the Lagrangian for the relevant fermionic interactions of all these LQs as
\begin{align}
{\cal L}_{\textsc{lq}}^{} & \,=\, \Big[ \texttt Y_{1,jy}^{\textsc{rr}}\, \overline{d_j^{\rm c}}e_y^{}S{}_1^{}
+ \texttt Y_{2,jy}^{\textsc{lr}}\, \overline{q_j^{}} e_y^{} S_2^{}
+ \texttt Y_{2,jy}^{\textsc{rl}}\, \overline{d_j^{}} \tilde S_2^{\textsc t\,} \varepsilon l_y^{}
+ \texttt Y_{3,jy}^{\textsc{ll}}\, \overline{q_j^{\rm c}}\, \varepsilon\, \tau_{\texttt I}^{} l_y^{} S_{3,\texttt I}^{}
\nonumber \\ & ~~~ ~~ +
\Big( \texttt Z_{1,jy}^{\textsc{ll}}\, \overline{q_j^{}} \gamma_\eta^{} l_y^{}
+ \texttt Z_{1,jy}^{\textsc{rr}}\, \overline{d_j^{}} \gamma_\eta^{} e_y^{} \Big) V_1^\eta
+ \texttt Z_{2,jy}^{\textsc{rl}}\, \overline{d_j^{\rm c}} \gamma_\eta^{} V_2^{\eta\textsc t} \varepsilon l_y^{}
+ \texttt Z_{2,jy}^{\textsc{lr}}\, \overline{q_j^{\rm c}} \gamma_\eta^{} \varepsilon V_2^\eta e_y^{}
\nonumber \\ & ~~~ ~~ +\,
\texttt Z_{3,jy}^{\textsc{ll}}\, \overline{q_j^{}} \gamma_\eta^{} \tau_{\texttt I}^{} l_y^{} V_{3,\texttt I}^\eta \Big] \,+\, \rm H.c. \,,
\end{align}
where the $\texttt Y_{jy}$ and $\texttt Z_{jy}$ are dimensionless free parameters which can be complex, the superscript c indicates charge conjugation, summation over $j,y,\texttt I$\, is implicit, and  \,$\varepsilon=i\tau_2^{}$.\,

From ${\cal L}_{\textsc{lq}}$, we can then derive LQ-mediated quark-lepton couplings at tree level which yield the operators in eq.\,(\ref{Lnpo}), with their coefficients being given by
\begin{align}
\frac{{\cal C}_1^{ijxy}}{\Lambda_{\textsc{np}}^2} & =
\frac{3\texttt Y_{3,ix}^{\textsc{ll}*} \texttt Y_{3,jy}^{\textsc{ll}}}{4 m_{S_3}^2}
- \frac{\texttt Z_{1,jx}^{\textsc{ll}*} \texttt Z_{1,iy}^{\textsc{ll}}}{2 m_{V_1}^2}
- \frac{3 \texttt Z_{3,jx}^{\textsc{ll}*} \texttt Z_{3,iy}^{\textsc{ll}}}{2 m_{V_3}^2} \,, ~~ &
\frac{{\cal C}_2^{ijxy}}{\Lambda_{\textsc{np}}^2} & =
\frac{\texttt Y_{3,ix}^{\textsc{ll}*} \texttt Y_{3,jy}^{\textsc{ll}}}{4 m_{S_3}^2}
- \frac{\texttt Z_{1,jx}^{\textsc{ll}*} \texttt Z_{1,iy}^{\textsc{ll}}}{2 m_{V_1}^2}
+ \frac{\texttt Z_{3,jx}^{\textsc{ll}*} \texttt Z_{3,iy}^{\textsc{ll}}}{2 m_{V_3}^2} \,,
\nonumber \\
\frac{{\cal C}_3^{ijxy}}{\Lambda_{\textsc{np}}^2} & =
\frac{\texttt Y_{1,ix}^{\textsc{rr}*} \texttt Y_{1,jy}^{\textsc{rr}}}{2m_{S_1}^2}
- \frac{\texttt Z_{1,jx}^{\textsc{rr}*} \texttt Z_{1,iy}^{\textsc{rr}}}{m_{V_1}^2} \,, &
\frac{{\cal C}_4^{ijxy}}{\Lambda_{\textsc{np}}^2} & =
\frac{-\tilde{\texttt Y}_{2,jx}^{\textsc{rl}*}\tilde{\texttt Y}_{2,iy}^{\textsc{rl}}}
{2m_{\tilde S_2}^2}
+ \frac{\texttt Z_{2,ix}^{\textsc{rl}*} \texttt Z_{2,jy}^{\textsc{rl}}}{m_{V_2}^2} \,,
\nonumber \\
\frac{{\cal C}_5^{ijxy}}{\Lambda_{\textsc{np}}^2} & =
\frac{-\texttt Y_{2,jx}^{\textsc{lr}*}\texttt Y_{2,iy}^{\textsc{lr}}}{2m_{S_2}^2}
+ \frac{\texttt Z_{2,ix}^{\textsc{lr}*} \texttt Z_{2,jy}^{\textsc{lr}}}{m_{V_2}^2} \,, &
\frac{{\cal C}_6^{ijxy}}{\Lambda_{\textsc{np}}^2} & =
\frac{2 \texttt Z_{1,yi}^{\textsc{ll}*} \texttt Z_{1,xj}^{\textsc{rr}}}{m_{V_1}^2}
- \frac{2 \texttt Z_{2,xi}^{\textsc{rl}*} \texttt Z_{2,yj}^{\textsc{lr}}}{m_{V_2}^2} \,.
\end{align}
Evidently ${\cal C}_{1,...,5}$ can all be affected by the scalar and vector LQs, but ${\cal C}_6$ only by the vector ones.

\section{Hadronic matrix elements and decay rates\label{drates}}

\subsection{Hyperon decays\label{hyperons}}

Our baryon decays of interest are \,$\mathfrak B\to\mathfrak B'e^\mp\mu^\pm$\, for
\,$\mathfrak{BB}'=\Lambda n,\Sigma^+p,\Xi^0\Lambda,\Xi^0\Sigma^0,\Xi^-\Sigma^-$,\, all involving spin-1/2 particles only,\footnote{We do not include \,$\Sigma^0\to n e^\mp \mu^\pm$\, because their branching fractions are expected to be comparatively much smaller due to the $\Sigma^0$ width being overwhelmingly dominated by the electromagnetic channel \,$\Sigma^0\to\Lambda\gamma$ \cite{Tanabashi:2018oca}.} and \,$\Omega^-\to\Xi^-e^\mp\mu^\pm$,\, where $\Omega^-$ is a spin-3/2 hyperon.
To determine their amplitudes, we need the baryonic matrix elements of \,$\overline{d}\big(\gamma^\eta,\gamma^\eta\gamma_5^{},1,\gamma_5^{}\big)s$, which can be estimated with the aid of chiral perturbation theory at leading order.
Their derivation from the chiral Lagrangian is sketched in appendix \ref{correspondences}.
For \,$\mathfrak B\to\mathfrak B'e^\mp\mu^\pm$,\, the results are
\begin{align} \label{<B'B>}
\big\langle\mathfrak B'\big|\overline{d}\gamma^\eta s\big|\mathfrak B\big\rangle & \,=\,
{\cal V}_{\mathfrak B'\mathfrak B}^{}\,\bar u_{\mathfrak B'}^{}\gamma^\eta u_{\mathfrak B}^{} \,,
~ & \big\langle{\mathfrak B'}\big|\overline{d}\gamma^\eta\gamma_5^{}s\big|{\mathfrak B}\big\rangle
& \,=\, \bar u_{\mathfrak B'}^{} \bigg( \gamma^\eta {\cal A}_{\mathfrak B'\mathfrak B}^{}
- \frac{{\cal P}_{\mathfrak B'\mathfrak B}}{B_0}\, \hat{\texttt Q}{}^\eta \bigg)
\gamma_5^{} u_{\mathfrak B}^{} \,, &
\nonumber \\
\big\langle{\mathfrak B}'\big|\overline{d}s\big|{\mathfrak B}\big\rangle & \,=\,
{\cal S}_{\mathfrak B'\mathfrak B}^{}\, \bar u_{\mathfrak B'}^{} u_{\mathfrak B}^{} \,, &
\big\langle{\mathfrak B'}\big|\overline{d} \gamma_5^{}s\big|{\mathfrak B}\big\rangle
& \,=\, {\cal P}_{\mathfrak B'\mathfrak B}^{}\,
\bar u_{\mathfrak B'}^{}\gamma_5^{}u_{\mathfrak B}^{} \,,
\end{align}
where ${\cal V}_{\mathfrak B'\mathfrak B}$ and ${\cal A}_{\mathfrak B'\mathfrak B}$ are constants, their values for the aforesaid $\mathfrak B'\mathfrak B$ pairs collected in table\,\,\ref{VA}, the $u$s are Dirac spinors, \,$\hat{\texttt Q}=p_{\mathfrak B}^{}-p_{\mathfrak B'}^{}$,\, with $p_X^{}$ denoting the four-momentum of $X$,
\begin{align}
{\cal S}_{\mathfrak B'\mathfrak B}^{} & \,=\, \frac{m_{\mathfrak B}^{}-m_{\mathfrak B'}^{}}
{m_s^{}-\hat m}\, {\cal V}_{\mathfrak B'\mathfrak B}^{} \,, &
{\cal P}_{\mathfrak B'\mathfrak B}^{} & \,=\, {\cal A}_{\mathfrak B'\mathfrak B\,}^{} B_0^{}~
\frac{m_{\mathfrak B'}^{}+m_{\mathfrak B}^{}}{m_K^2-\hat{\texttt Q}{}^2} \,,
\end{align}
and the other quantities are defined in appendix \ref{correspondences}.
In the \,$\Omega^-\to\Xi^-e^\mp\mu^\pm$\, case, we have
\begin{align} \label{<X-O->}
\langle\Xi^-\big|\overline{d}\gamma^\eta\gamma_5^{}s|\Omega^-\rangle & \,=\, {\cal C}\,
\bar u_{\Xi}^{} \bigg( u_{\Omega}^\eta + \frac{\tilde{\textsc q}{}^\eta\,
\tilde{\textsc q}{}_\kappa^{}}{m_K^2-\tilde{\textsc q}{}^2}\, u_\Omega^\kappa \bigg) , &
\langle\Xi^-|\overline{d}\gamma_5^{}s|\Omega^-\rangle & \,=\, \frac{B_0^{}\,{\cal C}\,
\tilde{\textsc q}_\kappa^{}}{\tilde{\textsc q}{}^2-m_K^2}\,\bar u_{\Xi}^{}u_\Omega^\kappa \,,
\end{align}
and \,$\langle\Xi^-|\overline{d}\gamma^\eta s|\Omega^-\rangle =
\langle\Xi^-|\overline{d} s|\Omega^-\rangle = 0$,\, where \,$\tilde{\textsc q}=p_{\Omega^-}^{}-p_{\Xi^-}^{}$\, and $u_\Omega^\eta$ is a Rarita-Schwinger spinor.

\begin{table}[h] \medskip \hspace*{\fill}
\begin{tabular}{|c||c|c|c|c|c|c|} \hline
$\mathfrak B'\mathfrak B$ & $n\Lambda$ & $p\Sigma^+$ & $\Lambda\Xi^0$ & $\Sigma^0\Xi^0$ &
$\Sigma^-\Xi^-\vphantom{\int_o^|}$ \\ \hline\hline
~${\cal V}_{\mathfrak B'\mathfrak B}^{}$~ & $-\sqrt{\tfrac{3}{2}}$ & $-1$ &
$\sqrt{\tfrac{3}{2}}^{\vphantom{|}}$ & $\tfrac{-1}{\sqrt2}_{\vphantom{\int}^{}}$ & $1$ \\ \hline
${\cal A}_{\mathfrak B'\mathfrak B}^{}$ &
~$\displaystyle \tfrac{-1}{\sqrt6}_{\vphantom{\int}}^{\vphantom{\int}}(D+3F)$~ & ~$D-F$~ &
~$\displaystyle \tfrac{-1}{\sqrt6}(D-3F)$~ & ~$\displaystyle \tfrac{-1}{\sqrt2}(D+F)$~ &
~$D+F\vphantom{\int_{\int_|}^{\int}}$~ \\ \hline
\end{tabular} \hspace*{\fill}
\caption{Values of ${\cal V}_{\mathfrak B'\mathfrak B}$ and ${\cal A}_{\mathfrak B'\mathfrak B}$ in eq.\,(\ref{<B'B>}) for \,$\mathfrak{BB}'=\Lambda n,\Sigma^+p,\Xi^0\Lambda,\Xi^0\Sigma^0,\Xi^-\Sigma^-$.\,
The parameters $D$ and $F$ are from the lowest-order chiral Lagrangian.} \label{VA}
\end{table}

In numerical work, to incorporate form-factor effects not taken into account in eq.\,(\ref{<B'B>}), we will modify \,${\cal V}_{\mathfrak B'\mathfrak B}$\, and \,${\cal A}_{\mathfrak B'\mathfrak B}$\, to \,$\big(1+2\hat{\texttt Q}{}^2/M_V^2\big){\cal V}_{\mathfrak B'\mathfrak B}$\, and
\,$\big(1+2\hat{\texttt Q}{}^2/M_A^2\big){\cal A}_{\mathfrak B'\mathfrak B}$,\, respectively, with \,\mbox{$M_V=0.97(4)$ GeV}\, and \,$M_A=1.25(15)$ GeV,\, following the commonly used parametrization in experimental analyses of semileptonic hyperon decays \cite{Bourquin:1981ba,Gaillard:1984ny,Hsueh:1988ar,Dworkin:1990dd,Batley:2006fc} and assuming isospin symmetry.
Analogously, as the $\tilde{\textsc q}{}^2$ range in \,$\Omega^-\to\Xi^-e^\mp\mu^\pm$\, is significantly larger than the $\hat{\texttt Q}{}^2$ ones, for this decay we will make the change \,${\cal C}\to{\cal C}/ \big(1-\tilde{\textsc q}{}^2/M_A^2\big)\raisebox{1pt}{$^2$}$.\,
With the central values of the input parameters, these modifications turn out to translate into increases of the decay rates ranging from a few percent to about 20\%.
The $M_{V,A}$ ranges quoted above lead to a rate uncertainty of under 3\% (6\%) in the spin-1/2 hyperon $(\Omega^-)$ case.\footnote{Our finding of 3\% is compatible with the values of 2\% or less estimated in the experimental analyses of spin-1/2 hyperon semileptonic decays \cite{Gaillard:1984ny,Hsueh:1988ar,Dworkin:1990dd,Batley:2006fc}.
It is also consistent with or smaller than estimates of uncertainties from higher-order corrections in chiral perturbation theory.}

With eq.\,(\ref{<B'B>}), we can express the amplitude for the spin-1/2 hyperon decay
\,$\mathfrak B\to\mathfrak B'\ell^-\ell^{\prime+}$\, induced by the interactions in
eq.\,(\ref{s->dem}) as
\begin{align}
{\cal M}_{\mathfrak B\to\mathfrak B'\ell\bar{\ell'}}^{} \,=\, &~
\bar u_{\mathfrak B'}^{} \gamma^\eta u_{\mathfrak B}^{}\, \bar u_\ell^{} \gamma_\eta^{} \Big[
V_{\mathfrak{BB}{}'\ell\ell'}^{} + \gamma_5^{} A_{\mathfrak{BB}{}'\ell\ell'}^{} \Big] v_{\ell'}^{}
\nonumber \\ & +\,
\bar u_{\mathfrak B'}^{} \gamma^\eta\gamma_5^{} u_{\mathfrak B}^{}\, \bar u_\ell^{}
\gamma_\eta^{} \Big[ \tilde V_{\mathfrak{BB}{}'\ell\ell'}^{}
+ \gamma_5^{} \tilde A_{\mathfrak{BB}{}'\ell\ell'}^{} \Big] v_{\ell'}^{}
\nonumber \\ & +\,
\bar u_{\mathfrak B'}^{}u_{\mathfrak B}^{}\, \bar u_\ell^{} \Big[ S_{\mathfrak{BB}{}'\ell\ell'}^{}
+\gamma_5^{} P_{\mathfrak{BB}{}'\ell\ell'}^{} \Big] v_{\ell'}^{} 
\nonumber \\ & +\,
\bar u_{\mathfrak B'}^{} \gamma_5^{}u_{\mathfrak B}^{}\, \bar u_\ell^{} \Big[ \tilde S_{\mathfrak{BB}{}'\ell\ell'}^{} + \gamma_5^{} \tilde P_{\mathfrak{BB}{}'\ell\ell'}^{} \Big] v_{\ell'}^{} \,,
\end{align}
where
\begin{align}
V_{\mathfrak{BB}{}'\ell\ell'}^{} & = {\cal V}_{\mathfrak B'\mathfrak B}^{}\,
\frac{{\texttt V}_{\ell\ell'}^{}}{\Lambda_{\textsc{np}}^2} \,, & 
A_{\mathfrak{BB}{}'\ell\ell'}^{} & = {\cal V}_{\mathfrak B'\mathfrak B}^{}\,
\frac{{\texttt A}_{\ell\ell'}^{}}{\Lambda_{\textsc{np}}^2} \,, 
\nonumber \\
S_{\mathfrak{BB}{}'\ell\ell'}^{} & = {\cal S}_{\mathfrak B'\mathfrak B}^{}\,
\frac{{\texttt S}_{\ell\ell'}^{}}{\Lambda_{\textsc{np}}^2} \,, & 
P_{\mathfrak{BB}{}'\ell\ell'}^{} & = {\cal S}_{\mathfrak B'\mathfrak B}^{}\,
\frac{{\texttt P}_{\ell\ell'}^{}}{\Lambda_{\textsc{np}}^2} \,, 
\nonumber \\
\tilde V_{\mathfrak{BB}{}'\ell\ell'}^{} & = {\cal A}_{\mathfrak B'\mathfrak B}^{}\,
\frac{\tilde{\textsc v}_{\ell\ell'}^{}}{\Lambda_{\textsc{np}}^2} \,, & 
\tilde A_{\mathfrak{BB}{}'\ell\ell'}^{} & = {\cal A}_{\mathfrak B'\mathfrak B}^{}\,
\frac{\tilde{\textsc a}_{\ell\ell'}^{}}{\Lambda_{\textsc{np}}^2} \,,  
\nonumber \\
\tilde S_{\mathfrak{BB}{}'\ell\ell'}^{} & = \frac{{\cal P}_{\mathfrak B'\mathfrak B}^{}}{\Lambda_{\textsc{np}}^2} \bigg[ \tilde{\textsc s}_{\ell\ell'}^{} - \frac{m_\ell^{}-m_{\ell'}^{}}{B_0} \tilde{\textsc v}_{\ell\ell'}^{} \bigg] , &
\tilde P_{\mathfrak{BB}{}'\ell\ell'}^{} & = \frac{{\cal P}_{\mathfrak B'\mathfrak B}^{}}{\Lambda_{\textsc{np}}^2} \bigg[ \tilde{\textsc p}_{\ell\ell'}^{} - \frac{m_\ell^{}+m_{\ell'}^{}}{B_0} \tilde{\textsc a}_{\ell\ell'}^{} \bigg] ,
\end{align}
Hereafter we neglect the electron mass.
Defining
\,$\Gamma_{X\to Y\ell^-\ell^{\prime+}}'\equiv d\Gamma_{X\to Y\ell^-\ell^{\prime+}}/d\hat s$\,
for the differential decay rate, we then arrive at
\begin{align} \label{G'B2B'em}
\Gamma_{\mathfrak B\to\mathfrak B'e^-\mu^+}' = &~
\frac{\beta^4 \lambda_{\mathfrak{BB}'}^{1/2}}{64\pi^3 m_{\mathfrak B}^3} \Bigg\{
\Bigg[ \frac{3-2\beta^2}{3}\lambda_{\mathfrak{BB}'}^{}+\hat m{}_-^2\hat s
+ \frac{m_\mu^2}{2} \big(\hat m{}_+^2+\hat m{}_-^2\big) \Bigg] \Big[
\big|V_{\mathfrak{BB}{}'e\mu}\big|^2 + \big|A_{\mathfrak{BB}{}'e\mu}\big|^2 \Big]
\nonumber \\ & \hspace{4em} \,+
\Bigg[ \frac{3-2\beta^2}{3}\lambda_{\mathfrak{BB}'}^{}+\hat m{}_+^2\hat s
+ \frac{m_\mu^2}{2} \big(\hat m{}_+^2+\hat m{}_-^2\big) \Bigg] \Big[
\big|\tilde V_{\mathfrak{BB}{}'e\mu}\big|^2 + \big|\tilde A_{\mathfrak{BB}{}'e\mu}\big|^2 \Big]
\nonumber \\ & \hspace{4em} \,+\,
m_\mu^{}\, {\rm Re} \Big[ \hat m{}_+^2\, {\texttt M}_-^{}\, \Big( A_{\mathfrak{BB}{}'e\mu}^*
P_{\mathfrak{BB}{}'e\mu}^{} - V_{\mathfrak{BB}{}'e\mu}^* S_{\mathfrak{BB}{}'e\mu}^{} \Big)
\nonumber \\ & \hspace{8em} ~-\,
\hat m{}_-^2\, {\texttt M}_+^{}\, \Big( \tilde A{}_{\mathfrak{BB}{}'e\mu}^*
\tilde P_{\mathfrak{BB}{}'e\mu}^{} - \tilde V{}_{\mathfrak{BB}{}'e\mu}^*
\tilde S_{\mathfrak{BB}{}'e\mu}^{} \Big) \Big]
\nonumber \\ & \hspace{4em} \,+\,
\hat m{}_+^2 \Big[ \big|S_{\mathfrak{BB}{}'e\mu}\big|^2
+ \big|P_{\mathfrak{BB}{}'e\mu}\big|^2 \Big] \frac{\hat s}{2}
+ \hat m{}_-^2 \Big[ \big|\tilde S_{\mathfrak{BB}{}'e\mu}\big|^2
+ \big|\tilde P_{\mathfrak{BB}{}'e\mu}\big|^2 \Big] \frac{\hat s}{2} \Bigg\} \,,
\end{align}
where
\begin{align}
\beta & \,=\, \sqrt{1-\frac{m_\mu^2}{\hat s}} \,, & \hat s & \,=\, \big(p_e^{}+p_\mu\big)^2 \,, &
\lambda_{XY}^{} & \,=\, m_X^4-2\big(m_Y^2+\hat s\big)m_X^2 + \big(m_Y^2-\hat s\big)\raisebox{1pt}{$^2$} \,,
\nonumber \\
\hat m{}_\pm^2 & \,=\, {\texttt M}_\pm^2-\hat s \,, &
{\texttt M}_\pm^{} & \,=\, m_{\mathfrak B}^{}\pm m_{\mathfrak B'}^{} \,.
\end{align}
Similarly, for the $\Omega^-$ decay, we find
\begin{align}
{\cal M}_{\Omega^-\to\Xi^-\ell\bar{\ell'}}^{} \,=\, \frac{\cal C}{\Lambda_{\textsc{np}}^2}
\bigg( g_{\kappa\varsigma}^{} + \frac{\hat p_\kappa^{}\,\hat p_\varsigma^{}}{\tilde{\textsc k}{}^2}
\bigg) \bar u_\Xi^{} u_\Omega^\kappa\, \bar u_\ell^{} \Bigg[ \gamma^\varsigma \big(
\tilde{\textsc v}_{\ell\ell'}^{} + \gamma_5^{} \tilde{\textsc a}_{\ell\ell'}^{} \big)
- \frac{B_0^{}\, \hat p^\varsigma}{m_K^2} \big( \tilde{\textsc s}_{\ell\ell'}^{}
+ \gamma_5^{} \tilde{\textsc p}_{\ell\ell'}^{} \big) \Bigg] v_{\ell'}^{} \,,
\end{align}
where \,$\hat p=p_e^{}+p_\mu^{}=\tilde{\textsc q}$\, and \,$\tilde{\textsc k}{}^2=m_K^2-\hat s$.\,
Hence
\begin{align} \label{G'O2Xem}
\Gamma_{\Omega^-\to\Xi^-e^-\mu^+}' = \frac{\beta^4 \lambda_{\Omega^-\Xi^-}^{1/2} {\cal C}^2 \tilde{\texttt M}{}^2}{384\pi^3 \Lambda_{\textsc{np}}^4 m_{\Omega^-}^3} & \Bigg\{ \! \Bigg[ 3\hat s-\beta^2 \hat s
+ \frac{\lambda_{\Omega^-\Xi^-}^{}}{m_{\Omega^-}^2} \Bigg( \frac{1}{2} - \frac{\beta^2}{3}
+ \frac{2 \tilde{\textsc k}{}^2+\hat s}{4\, \tilde{\textsc k}{}^4} m_\mu^2 \Bigg)
\Bigg] \Big[ |\tilde{\textsc v}_{e\mu}|^2+|\tilde{\textsc a}_{e\mu}|^2 \Big]
\nonumber \\ & ~ -_{\;}\!
\frac{\lambda_{\Omega^-\Xi^-}^{} B_{0\,}^{} m_{\mu\,}^{} m_K^2}{2\, \tilde{\textsc k}{}^{4\,} m_{\Omega^-}^2}\,
{\rm Re} \Big( \tilde{\textsc a}_{e\mu}^* \tilde{\textsc p}_{e\mu}^{} - \tilde{\textsc v}_{e\mu}^*
\tilde{\textsc s}_{e\mu}^{} \Big)
\nonumber \\ & ~ +_{\;}\!
\frac{\lambda_{\Omega^-\Xi^-}^{} B_{0\,}^2 \hat s}{4\, \tilde{\textsc k}{}^{4\,} m_{\Omega^-}^2}
\Big(\big|\tilde{\textsc s}_{e\mu}\big|^2+\big|\tilde{\textsc p}_{e\mu}\big|^2\Big) \Bigg\} \,,
\end{align}
where \,$\tilde{\texttt M}{}^2=(m_{\Omega^-}+m_{\Xi^-})^2-\hat s$.\,
Thus, in our approximation of the \,$\Omega^-\to\Xi^-$\, matrix elements, \,$\Omega^-\to\Xi^-e^-\mu^+$\, is not sensitive to the untilded couplings $\texttt V_{e\mu}$ and $\texttt A_{e\mu}$ but indirectly still probes $\texttt S_{e\mu}$ and $\texttt P_{e\mu}$ in light of eq.\,(\ref{sp}).
The differential rates of the $\mu^-e^+$ modes are obtainable from their $e^-\mu^+$ counterparts by interchanging $e$ and $\mu$ in the subscripts of $\texttt A_{e\mu}$, $\texttt S_{e\mu}$, $\texttt P_{e\mu}$, and $\tilde{\textsc a}_{e\mu}$ as well as applying \,$\texttt V_{e\mu}\to-\texttt V_{\mu e}$\, and \,$\tilde{\textsc v}_{e\mu}\to-\tilde{\textsc v}_{\mu e}$.\,

\subsection{Kaon decays\label{kaons}}

For \,$K_{L,S}\to e^\mp\mu^\pm$\, the pertinent hadronic matrix elements are
\begin{align} \label{<vacK>}
\langle0|\overline{d}\gamma^\eta\gamma_5^{}s|\bar K^0\rangle & =
\langle0|\overline{s}\gamma^\eta\gamma_5^{}d|K^0\rangle =\, -i f_K^{} p_K^\eta \,, &
\nonumber \\
\big\langle0\big|\overline{d}\gamma_5^{}s\big|\bar K^0\big\rangle & =
\big\langle0\big|\overline{s}\gamma_5^{}d\big|K^0\big\rangle =\, i B_0^{}f_K^{} \,,
\end{align}
with $f_K^{}$ being the kaon decay constant, while for \,$K\to\pi e^\mp\mu^\pm$\,
\begin{align}
\big\langle\pi^-\big|\bar d\gamma^\eta s\big|K^-\big\rangle & \,=\,
-\langle\pi^+|\bar s\gamma^\eta d|K^+\rangle \,=\,
\big(p_K^\eta+p_\pi^\eta\big) f_+^{} \,+\, \big(f_0^{}-f_+^{}\big) q_{K\pi}^\eta\,
\frac{m_K^2-m_\pi^2}{q_{K\pi}^2} \,, &
\nonumber \\ \label{<pi-K->}
\big\langle\pi^-\big|\bar d s\big|K^-\big\rangle & \,=\,
+\big\langle\pi^+\big|\bar s d\big|K^+\big\rangle \,=\, B_0^{} f_0^{} \,, ~~~ ~~~~ ~~~
q_{\texttt{XY}}^{} \,=\, p_{\texttt X}^{}-p_{\texttt Y}^{} \,,
\end{align}
where $f_{+,0}^{}$ represent form factors which are functions of $q_{K\pi}^2$.
Additional required matrix elements are
\,$\big\langle\pi^0\big|\bar d(\gamma^\eta,1)s\big|\bar K^0\big\rangle =
\big\langle\pi^0\big|\bar s(-\gamma^\eta,1)d\big|K^0\big\rangle =
-\big\langle\pi^-\big|\bar d(\gamma^\eta,1)s\big|K^-\big\rangle/\sqrt2$\,
under the assumption of isospin symmetry, which also implies
\,$\big\langle\pi^-\big|\bar d\gamma^\eta s\big|K^-\big\rangle=
\big\langle\pi^+\big|\bar u\gamma^\eta s\big|\bar K^0\big\rangle$.\,
This allows us to adopt
$f_{+,0}^{}=\texttt f_+^{}(0)\big(1+\lambda_{+,0}^{}\,q_{K\pi}^2/m_{\pi^+}^2\big)$\,
with \,$\lambda_+^{}=0.0271(10)$\, and \,$\lambda_0^{}=0.0142(23)$\, from
\,$K_L\to\pi^+\mu^-\nu$ measurements \cite{Tanabashi:2018oca} as well as \,$\texttt f_+^{}(0)=0.9681(23)$\,
from lattice computations \cite{Charles:2015gya}.\footnote{Online updates are available at
http://ckmfitter.in2p3.fr.}
It is simple to check that the baryonic and mesonic matrix elements detailed above satisfy the free quark relations
\,$\langle\texttt Y|\overline{d}\gamma_\kappa^{}s|\texttt X\rangle
q_{\texttt{XY}}^\kappa = (m_s^{}-m_d)\langle\texttt Y|\overline{d}s|\texttt X\rangle$\,
and
\,$\langle\texttt Y|\overline{d}\gamma_\kappa^{}\gamma_5^{}s|\texttt X\rangle q_{\texttt{YX}}^\kappa
= (m_s^{}+m_d)\langle\texttt Y|\overline{d}\gamma_5^{}s|\texttt X\rangle$.

The amplitude for \,${\cal K}\to\ell^-\ell^{\prime+}$\, has the form
\begin{equation} \label{MK2ll'}
{\cal M}_{{\cal K}\to\ell\bar{\ell'}}^{} \,=\, i\, \bar u_\ell^{} \Big(
S_{{\cal K}\ell\ell'}^{} \,+\, \gamma_5^{} P_{{\cal K}\ell\ell'}^{} \Big) v_{\ell'}^{} \,.
\end{equation}
After the absolute square of the amplitude is summed over the final spins, there is no
interference between the $S$ and $P$ terms.
This leads to the decay rates
\begin{equation} \label{GK2em}
\Gamma_{K\!_{L,S}^{}\to e^-\mu^+} ^{} \,=\, \Gamma_{K\!_{L,S}^{}\to\mu^-e^+}^{} \,=\,
\frac{\big(m_{K^0}^2-m_\mu^2\big)\raisebox{1pt}{$^2$}}{8\pi\, m_{K^0}^3}
\Big( \big| S_{K\!_{L,S\,}^{}e\mu} \big| \raisebox{1pt}{$^2$}
+ \big| P_{K\!_{L,S\,}^{}e\mu} \big| \raisebox{1pt}{$^2$} \Big) .
\end{equation}
The expressions for $S_{K\!_{L,S\,}e\mu}$ and $P_{K\!_{L,S\,}e\mu}$ have been relegated to appendix \ref{Kformulas}.

For \,${\cal K}\to\pi\ell^-\ell^{\prime+}$,\, the amplitude is
\begin{equation} \label{MK2pll'}
{\cal M}_{{\cal K}\to\pi\ell\bar{\ell'}}^{} \,=\, \bar u_\ell^{} \Big(
S_{{\cal K}\pi\ell\ell'}^{} \,+\, P_{{\cal K}\pi\ell\ell'\,}^{} \gamma_5^{} \Big) v_{\ell'}^{} \,.
\end{equation}
The resulting differential decay rates of \,$K^\mp\to\pi^\mp\mu^\mp e^\pm$,\, $K^\mp\to\pi^\mp e^\mp\mu^\pm$,\, and \,$K_{L,S}\to\pi^0\mu^\mp e^\pm$\, are collected in appendix \ref{Kformulas} as well.

\subsection{Other modes\label{othermodes}}

As mentioned earlier, there are other modes that can be influenced by $Q_k^{e\mu,\mu e}$ in eq.\,\,(\ref{Lnp}).
The relevant observables are affected as follows.

\begin{itemize}

\item Modes with two neutrinos

$K^+ \to \pi^+ \nu \bar \nu$\, and \,$K_L \to \pi^0 \nu \bar \nu$.\,
The additions to their SM branching fractions are generated by the $(\bar d s)(\bar\nu_e\nu_\mu)$ interaction listed in table \ref{tabFR}, plus its $\bar\nu_\mu\nu_e$ counterpart, and can be read off eqs.\,\,(9) and (10) in ref.\,\cite{He:2018uey} to be
\begin{align}
\Delta{\cal B}_{K^+}^{} & \,=\, \frac{\tilde\kappa_+^{}}{3}\left(\left|W_{e\mu}\right|^2+\left|W_{\mu e}\right|^2\right) , &
\Delta{\cal B}_{K_L}^{} & \,=\, \frac{\kappa_L^{}}{12} \big|W_{e\mu}-W_{\mu e}^*\big|^2 \,, &
\end{align}
where the prefactors are \,$\tilde\kappa_+^{} = 5.17\times10^{-11}$\, and \,$\kappa_L^{}=2.23\times 10^{-10}$ \cite{Buras:2004uu} and
\begin{align} \label{Wll'}
W_{\ell\ell'}^{} & \,\simeq\, 9700 \bigg(\frac{1\rm~TeV}{\Lambda_{\textsc{np}}}\bigg)\raisebox{7pt}{$\!^2$} \left(c_1^{\ell\ell'}-c_2^{\ell\ell'}+c_4^{\ell\ell'}\right) . &
\end{align}
Values of \,$|W_{\ell\ell'}|={\cal O}(1)$\, are currently allowed.

\item
The most important modification to the leptonic decay \,$M^+\to\ell^+\nu$\, of a pseudoscalar meson \,$M^+\sim\textsc u\bar{\textsc d}$\, ($\textsc u=u,c$\, and \,$\textsc d=d,s$) is from NP with LFV induced by (pseudo)scalar operators which are not helicity suppressed.
In our case, they are of the form
\begin{align}
{\cal L} & \,=\, \frac{-\textsl{\texttt C}_M^{}}{2\Lambda_{\textsc{np}}^2}\;
\overline{\textsc d}\gamma_5^{}\textsc u\; \overline{\nu_{\ell'}^{}} P_R^{} \ell \,.
\end{align}
This yields the biggest impact if \,$\ell=e$,\, in which case the SM rate is helicity suppressed the most.
With \,$\langle0|\overline{\textsc d}\gamma_5^{}\textsc u|M^+\rangle=i f_M^{}m_M^2/(m_{\textsc u}+m_{\textsc d})$\, and the $M$ decay constant $f_M^{}$, the modification to the rate is then
\begin{align} \label{DGM2enu}
\Delta\Gamma_{M^+\to e^+\nu}^{} & \,=\, \frac{|\textsl{\texttt C}_M|^2\, f_M^2\, m_M^5}
{64\pi \Lambda^4_{\textsc{np}\,}(m_{\textsc u}+m_{\textsc d})^2} \,,
\end{align}
analogously to eq.\,(\ref{GK2em}), the lepton masses having been ignored.
Note that there is no interference with the SM contribution as the neutrino is of the wrong flavor \cite{Valencia:1994cj}.
From the Feynman rules in appendix \ref{feynrules}, we infer
\begin{align} \label{CM}
\textsl{\texttt C}_\pi^{}   & \,=\, c_6^{\mu e\,} V_{us}^* \,, &
\textsl{\texttt C}_K^{}     & \,=\, c_{6\prime}^{e\mu*\,} V_{ud}^* \,, &
\textsl{\texttt C}_D^{}     & \,=\, c_6^{\mu e\,} V_{cs}^* \,, &
\textsl{\texttt C}_{D_s}^{} & \,=\, c_{6\prime}^{e\mu*\,} V_{cd}^* \,.
\end{align}
\item
$\mu\to e$ conversion in nuclei and \,$\pi^0,D^0\to e^\mp\mu^\pm$.\,
These arise from some of the operators responsible for \,$K_L\to e^\mp\mu^\pm$\, discussed above but,  according to appendix \ref{feynrules}, are not affected by the scalar operators $Q_{6,6\prime}^{\ell\ell'}$.
From the general formulas in ref.\,\cite{Kitano:2002mt}, we find the rate of \,$\mu^-\to e^-$\, conversion in nucleus $\cal N$ to be
\begin{align} \label{Bm2e}
{\cal B}(\mu^-{\cal N}\to e^-{\cal N}) & \,=\, \frac{m_\mu^5\,|V_{ud}V_{us}|^2}{\omega_{\rm capt}^{\cal N} \Lambda^4_{\textsc{np}}} \Big( \big| c_1^{e\mu}-c_2^{e\mu} \big|^2 + \big| c_5^{e\mu} \big|^2 \Big) \Big[ 2V_{\cal N}^{(p)} + V_{\cal N}^{(n)} \Big]^2 \,, &
\end{align}
where $V_{\cal N}^{(p,n)}$ are dimensionless integrals representing the overlap of $e$ and $\mu$ wave-functions for $\cal N$ and incorporating appropriate proton ($p$) and neutron ($n$) densities, and $\omega_{\rm capt}^{\cal N}$ is the rate of $\mu$ capture in $\cal N$.
For the meson decays, we obtain
\begin{align} \label{D2em}
\Gamma_{\pi^0\to e^\mp\mu^\pm}^{} & \,=\, \frac{f_\pi^2\, \big(m_{\pi^0}^2-m_\mu^2\big)\raisebox{1pt}{$^2$}
m_\mu^2\, |V_{ud}V_{us}|^2}{128\pi \Lambda_{\textsc{np}}^4 m_{\pi^0}^3} \Big[
\big| c_1^{e\mu}-c_2^{e\mu} \big|^2 + \big| c_5^{e\mu} \big|^2 + (e\leftrightarrow\mu) \Big] \,,
\nonumber \\
\Gamma_{D^0\to e^\mp\mu^\pm}^{} & \,=\, \frac{f_D^2\, \big(m_{D^0}^2-m_\mu^2\big)\raisebox{1pt}{$^2$}
m_\mu^2\, |V_{ud}V_{cs}|^2}{64\pi \Lambda_{\textsc{np}}^4 m_{D^0}^3} \Big[
\big| c_1^{e\mu}-c_2^{e\mu} \big|^2 + \big| c_5^{e\mu} \big|^2 + (e\leftrightarrow\mu) \Big] \,. &
\end{align}

\end{itemize}

\section{Numerical results\label{numeric}}

\subsection{Hyperon and kaon constraints\label{HKconstr}}

Integrating $\Gamma_{\mathfrak B\to\mathfrak B'e^-\mu^+}'$ over \,$m_\mu^2\le\hat s\le(m_{\mathfrak B}-m_{\mathfrak B'})^2$,\, we arrive at the branching fractions
\begin{align} \label{BB2B'em}
{\cal B}\big(\Lambda\to ne^-\mu^+\big) \,=\, & \Big[
0.73\, \big(|\texttt V_{e\mu}|^2+|\texttt A_{e\mu}|^2\big)
+ 1.7\, \big( |\texttt S_{e\mu}|^2 + |\texttt P_{e\mu}|^2 \big) + 1.8\; {\rm Re} \big(
\texttt A_{e\mu}^*\texttt P_{e\mu}^{} - \texttt V_{e\mu}^*\texttt S_{e\mu}^{} \big)
\nonumber \\ & \,+\, 1.1\, \big(|\tilde{\textsc v}_{e\mu}|^2+|\tilde{\textsc a}_{e\mu}|^2\big)
+ 0.21\, \big( |\tilde{\textsc s}_{e\mu}^{}|^2 + |\tilde{\textsc p}_{e\mu}|^2 \big)
\nonumber \\ & \,-\, 0.27\; {\rm Re} \big( \tilde{\textsc a}_{e\mu}^* \tilde{\textsc p}_{e\mu}^{}
- \tilde{\textsc v}_{e\mu}^* \tilde{\textsc s}_{e\mu}^{} \big) \Big]
\frac{10^7\rm\,GeV^4}{\Lambda_{\textsc{np}}^4} \,,
\end{align}
\begin{align}
{\cal B}\big(\Sigma^+\to pe^-\mu^+\big) \,=\, & \Big[
2.3\, \big(|\texttt V_{e\mu}|^2+|\texttt A_{e\mu}|^2\big)
+ 11\, \big(|\texttt S_{e\mu}|^2+|\texttt P_{e\mu}|^2\big) + 6.9\; {\rm Re} \big( \texttt A_{e\mu}^*
\texttt P_{e\mu}^{} - \texttt V_{e\mu}^*\texttt S_{e\mu}^{} \big)
\nonumber \\ & \,+\, 0.82\, \big(|\tilde{\textsc v}_{e\mu}|^2+|\tilde{\textsc a}_{e\mu}|^2\big)
+ 0.49\, \big( |\tilde{\textsc s}_{e\mu}|^2 + |\tilde{\textsc p}_{e\mu}|^2 \big)
\nonumber \\ & \,-\, 0.37\; {\rm Re} \big( \tilde{\textsc a}_{e\mu}^* \tilde{\textsc p}_{e\mu}^{}
- \tilde{\textsc v}_{e\mu}^* \tilde{\textsc s}_{e\mu}^{} \big) \Big]
\displaystyle\frac{10^7\rm\,GeV^4}{\Lambda_{\textsc{np}}^4} \,,
\end{align}
\begin{align}
{\cal B}\big(\Xi^0\to\Lambda e^-\mu^+\big) \,=\, & \Big[
2.4\, \big(|\texttt V_{e\mu}|^2+|\texttt A_{e\mu}|^2\big) + 7.5\, \big( |\texttt S_{e\mu}|^2
+ |\texttt P_{e\mu}|^2 \big) + 6.5\; {\rm Re} \big( \texttt A_{e\mu}^* \texttt P_{e\mu}^{}
- \texttt V_{e\mu}^*\texttt S_{e\mu}^{} \big)
\nonumber \\ & \,+\, 0.25\, \big(|\tilde{\textsc v}_{e\mu}|^2+|\tilde{\textsc a}_{e\mu}|^2\big)
+ 0.07\, \big( |\tilde{\textsc s}_{e\mu}^{}|^2 + |\tilde{\textsc p}_{e\mu}|^2 \big)
\nonumber \\ & \,-\, 0.08\; {\rm Re} \big( \tilde{\textsc a}_{e\mu}^* \tilde{\textsc p}_{e\mu}^{}
- \tilde{\textsc v}_{e\mu}^* \tilde{\textsc s}_{e\mu}^{} \big) \Big]
\frac{10^7\rm\,GeV^4}{\Lambda_{\textsc{np}}^4} \,.
\end{align}
Compared to these results, the corresponding numerical factors in ${\cal B}\big(\Xi^{0,-}\to\Sigma^{0,-}e^\pm\mu^\mp\big)$ turn out to be roughly at least two orders of magnitude lower, partly due to smaller phase space, and hence are not shown.
On the other hand, the $\Omega^-$ decay having comparatively greater phase space, its numbers are bigger by an order of magnitude or more,
\begin{align} \label{BO2Xem}
{\cal B}\big(\Omega^-\to\Xi^-e^-\mu^+\big) \,=\, & \Big[ 5.6\, \big( |\tilde{\textsc v}_{e\mu}|^2
+ |\tilde{\textsc a}_{e\mu}|^2\big) + 8.5\, \big( |\tilde{\textsc s}_{e\mu}^{}|^2
+ |\tilde{\textsc p}_{e\mu}|^2 \big)
\nonumber \\ & \,-\, 3.6\; {\rm Re} \big( \tilde{\textsc a}_{e\mu}^*
\tilde{\textsc p}_{e\mu}^{} - \tilde{\textsc v}_{e\mu}^* \tilde{\textsc s}_{e\mu}^{} \big) \Big]
\frac{10^8\rm\,GeV^4}{\Lambda_{\textsc{np}}^4} \,.
\end{align}
All the results in eqs.\,\,(\ref{BB2B'em})-(\ref{BO2Xem}) have included the form factors mentioned in subsection \ref{hyperons}.


For the two-body kaon decays, we calculate the branching fractions to be
\begin{align} \label{BK2em}
{\cal B}\big(K_L^{} & \to e^\pm\mu^\mp\big) \,=\, \tau_{K_L}^{} \big(
\Gamma_{K_L\to e^-\mu^+} + \Gamma_{K_L\to\mu^-e^+} \big)
\nonumber \\ & =\, 3.8\, \Big[ \big| \tilde{\textsc v}_{e\mu}^{} + \tilde{\textsc v}{}_{\mu e}^*
+ 19 \big( \tilde{\textsc s}_{e\mu}^{}-\tilde{\textsc s}{}_{\mu e}^* \big) \big|^2
+ \big| \tilde{\textsc a}_{e\mu}^{} + \tilde{\textsc a}{}_{\mu e}^* - 19 \big(
\tilde{\textsc p}_{e\mu}^{} + \tilde{\textsc p}{}_{\mu e}^* \big) \big|^2 \Big]
\frac{10^{11}\rm\,GeV^4}{\Lambda_{\textsc{np}}^4} \,,
\end{align}
\begin{align}
{\cal B}\big(K_S^{} & \to e^\pm\mu^\mp\big) \,=\, \tau_{K_S}^{} \big(
\Gamma_{K_S\to e^-\mu^+} + \Gamma_{K_S\to\mu^-e^+} \big)
\nonumber \\ & =\, 6.6\, \Big[ \big| \tilde{\textsc v}_{e\mu}^{} - \tilde{\textsc v}{}_{\mu e}^*
+ 19 \big( \tilde{\textsc s}_{e\mu}^{} + \tilde{\textsc s}{}_{\mu e}^* \big) \big|^2
+ \big| \tilde{\textsc a}_{e\mu}^{} - \tilde{\textsc a}{}_{\mu e}^* - 19 \big(
\tilde{\textsc p}_{e\mu}^{} - \tilde{\textsc p}{}_{\mu e}^* \big) \big|^2 \Big]
\frac{10^8\rm\,GeV^4}{\Lambda_{\textsc{np}}^4} \,,
\end{align}
having employed the central value of \,$f_K^{}=155.6(4)$\,MeV \cite{Tanabashi:2018oca}.
For \,$K\to\pi e^\mp\mu^\pm$,\, integrating their differential rates in appendix \ref{Kformulas} over
\,$m_\mu^2\le\hat s\le(m_K-m_\pi)^2$,\, we obtain
\begin{align}
{\cal B}\big(K_L^{} & \to\pi^0e^\pm\mu^\mp\big) \,=\, \tau_{K_L}^{}\, \big(
\Gamma_{K_L\to\pi^0e^-\mu^+} + \Gamma_{K_L\to\pi^0\mu^-e^+} \big)
\nonumber \\ & =\, 2.0\, \Big\{ \big|\texttt V_{e\mu}^*-\texttt V_{\mu e}^{} \big|\raisebox{1pt}{$^2$}
+ \big|\texttt A_{e\mu}^*-\texttt A_{\mu e}^{} \big|\raisebox{1pt}{$^2$}
+ 10\, \Big( \big|\texttt S_{e\mu}^*+\texttt S_{\mu e}^{}\big|^2
+ \big|\texttt P_{e\mu}^*-\texttt P_{\mu e}^{} \big|^2 \Big)
\nonumber \\ & ~~~~~ ~~~~ +\, 3.5\; {\rm Re} \Big[ \big( \texttt A_{e\mu}^{} - \texttt A_{\mu e}^* \big) \big( \texttt P_{e\mu}^* - \texttt P_{\mu e}^{} \big) - \big( \texttt V_{e\mu}^{} - \texttt V_{\mu e}^*
\big) \big( \texttt S_{e\mu}^* + \texttt S_{\mu e}^{} \big) \Big] \Big\}
\frac{10^{10}\rm\,GeV^4}{\Lambda_{\textsc{np}}^4} \,,
\end{align}
\begin{align} \label{B'KS}
{\cal B}\big(K_S^{} & \to\pi^0e^\pm\mu^\mp\big) \,=\, \tau_{K_S}^{} \big(
\Gamma_{K_S\to\pi^0e^-\mu^+} + \Gamma_{K_S\to\pi^0\mu^-e^+} \big)
\nonumber \\ & =\, 3.5\, \Big\{ \big|\texttt V_{e\mu}^*+\texttt V_{\mu e}\big|\raisebox{1pt}{$^2$}
+ \big|\texttt A_{e\mu}^*+\texttt A_{\mu e}\big|\raisebox{1pt}{$^2$}
+ 10\, \Big( \big|\texttt S_{e\mu}^*-\texttt S_{\mu e}\big|^2
+ \big|\texttt P_{e\mu}^*+\texttt P_{\mu e}\big|^2 \Big)
\nonumber \\ & ~~~~~ ~~~~ +\, 3.5\; {\rm Re} \Big[ \big( \texttt A_{e\mu} + \texttt A_{\mu e}^* \big) \big( \texttt P_{e\mu}^* + \texttt P_{\mu e} \big) - \big(\texttt V_{e\mu} + \texttt V_{\mu e}^*
\big) \big( \texttt S_{e\mu}^* - \texttt S_{\mu e} \big) \Big] \Big\}
\frac{10^7\rm\,GeV^4}{\Lambda_{\textsc{np}}^4} \,,
\end{align}
\begin{align} \label{B'K-}
{\cal B}\big(K^+\to\pi^+e^-\mu^+\big) \,=\, 8.7\, & \Big[
|\texttt V_{\mu e}|^2 + |\texttt A_{\mu e}|^2 + 10\, \big(
|\texttt S_{\mu e}|^2 + |\texttt P_{\mu e}|^2 \big)
\nonumber \\ & +\, 3.6\; {\rm Re} \big( \texttt A_{\mu e}^*
\texttt P_{\mu e}^{} + \texttt V_{\mu e}^* \texttt S_{\mu e}^{} \big) \Big]
\frac{10^9\rm\,GeV^4}{\Lambda_{\textsc{np}}^4} \,, \hspace{7em}
\end{align}
\begin{align}
{\cal B}\big(K^+\to\pi^+\mu^-e^+\big) \,=\, 8.7\, & \Big[
|\texttt V_{e\mu}|^2+|\texttt A_{e\mu}|^2 + 10\, \big(
|\texttt S_{e\mu}|^2 + |\texttt P_{e\mu}|^2 \big)
\nonumber \\ & +\, 3.6\; {\rm Re} \big( \texttt A_{e\mu}^*
\texttt P_{e\mu}^{} - \texttt V_{e\mu}^*\texttt S_{e\mu}^{} \big) \Big]
\frac{10^9\rm\,GeV^4}{\Lambda_{\textsc{np}}^4} \,. \hspace{7em}
\end{align}
We see that \,$K\to e^\pm\mu^\mp$ $\big(K\to\pi e^\pm\mu^\mp\big)$ are not sensitive to
$\texttt V_{\ell\ell'}$ and $\texttt A_{\ell\ell'}$ $\big(\tilde{\textsc v}_{\ell\ell'}$ and $\tilde{\textsc a}_{\ell\ell'}\big)$ but can still probe $\texttt S_{\ell\ell'}$ and $\texttt P_{\ell\ell'}$ $\big(\tilde{\textsc s}_{\ell\ell'}$ and $\tilde{\textsc p}_{\ell\ell'}\big)$ in light of eq.\,(\ref{sp}).

Currently there is not much empirical information on the lepton-flavor-violating decays of strange hadrons.
The only data available are the limits \cite{Tanabashi:2018oca}
\begin{align} \label{Klimits}
{\cal B}\big(K_L^{}\to e^\pm\mu^\mp\big) & \,<\, 4.7\times10^{-12} \,,
\nonumber \\
{\cal B}\big(K_L^{}\to\pi^0e^\pm\mu^\mp\big) & \,<\, 7.6\times10^{-11} \,,
\nonumber \\
{\cal B}\big(K^+\to\pi^+e^-\mu^+\big) & \,<\, 1.3\times10^{-11} \,,
\nonumber \\
{\cal B}\big(K^+\to\pi^+\mu^-e^+\big) & \,<\, 5.2\times10^{-10} \,,
\end{align}
and \,${\cal B}\big(K_L^{}\to\pi^0\pi^0e^\pm\mu^\mp\big)<1.7\times10^{-10}$,\, all at 90\% confidence level.
We will ignore the bound from $K_L^{}\to\pi^0\pi^0e^\pm\mu^\mp$ as it has smaller phase space than the other modes and probes the same couplings as \,$K_L\to e^\pm\mu^\mp$.\,
The numbers in eq.\,(\ref{Klimits}) and the corresponding formulas in eqs.\,(\ref{BK2em})-(\ref{B'K-}) translate, respectively, into the upper limits
\begin{align} \label{KL->em}
\Big[ \big| \tilde{\textsc v}_{e\mu}^{} + \tilde{\textsc v}{}_{\mu e}^*
+ 19 \big( \tilde{\textsc s}_{e\mu}^{}-\tilde{\textsc s}{}_{\mu e}^* \big) \big|^2
+ \big| \tilde{\textsc a}_{e\mu}^{} + \tilde{\textsc a}{}_{\mu e}^* - 19 \big(
\tilde{\textsc p}_{e\mu}^{} + \tilde{\textsc p}{}_{\mu e}^* \big) \big|^2 \Big]
\frac{10^{23}\rm\,GeV^4}{\Lambda_{\textsc{np}}^4} & \,<\, 1.2 \,, &
\end{align}
\begin{align} \label{KL->piem}
& \Big\{ \big|\texttt V_{e\mu}^*-\texttt V_{\mu e}\big|\raisebox{1pt}{$^2$}
+ \big|\texttt A_{e\mu}^*-\texttt A_{\mu e}\big|\raisebox{1pt}{$^2$}
+ 10 \Big( \big|\texttt S_{e\mu}^*+\texttt S_{\mu e}\big|^2
+ \big|\texttt P_{e\mu}^*-\texttt P_{\mu e}\big|^2 \Big)
\nonumber \\ & \,+\,
3.5\; {\rm Re} \big[ \big( \texttt A_{e\mu}^{} - \texttt A_{\mu e}^* \big) \big(
\texttt P_{e\mu}^* - \texttt P_{\mu e} \big) - \big( \texttt V_{e\mu} - \texttt V_{\mu e}^*
\big) \big( \texttt S_{e\mu}^* + \texttt S_{\mu e} \big) \big] \Big\}
\frac{10^{21}\rm\,GeV^4}{\Lambda_{\textsc{np}}^4}
\,<\, 3.8 \,,
\end{align}
\begin{align} \label{K+->pi+e-m+}
\Big[ |\texttt V_{\mu e}|^2 + |\texttt A_{\mu e}|^2
+ 10\, \big( |\texttt S_{\mu e}|^2 + |\texttt P_{\mu e}|^2 \big) + 3.6\; {\rm Re} \big(
\texttt A_{\mu e}^*\texttt P_{\mu e}^{} + \texttt V_{\mu e}^* \texttt S_{\mu e}^{} \big)
\Big] \frac{10^{21}\rm\,GeV^4}{\Lambda_{\textsc{np}}^4}  \,<\, 1.5 \,, &
\end{align}
\begin{align} \label{K+->pi+m-e+}
\Big[ |\texttt V_{e\mu}|^2+|\texttt A_{e\mu}|^2
+ 10\, \big( |\texttt S_{e\mu}|^2 + |\texttt P_{e\mu}|^2 \big) + 3.6\; {\rm Re} \big(
\texttt A_{e\mu}^*\texttt P_{e\mu}^{} - \texttt V_{e\mu}^*\texttt S_{e\mu}^{} \big)
\Big] \frac{10^{20}\rm\,GeV^4}{\Lambda_{\textsc{np}}^4}  \,<\, 6.0 \,. &
\end{align}

\begin{figure}[t!] \hspace*{\fill}%
\includegraphics[width=16em]{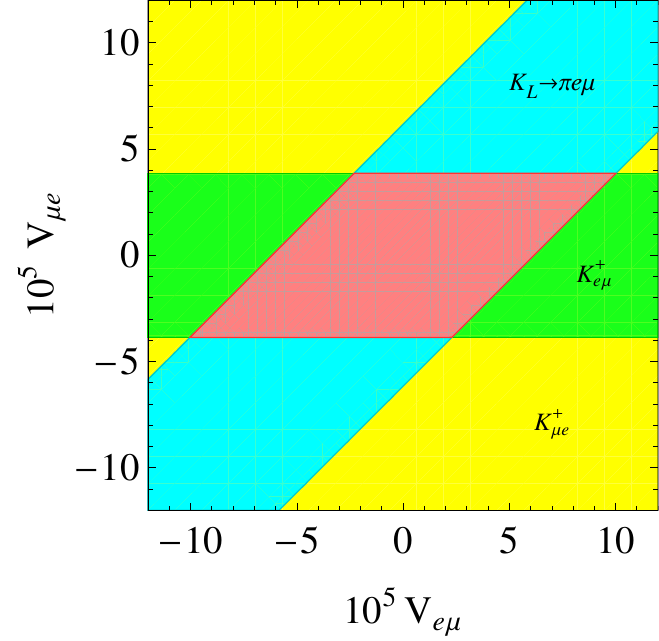} ~ ~ ~ ~
\includegraphics[width=16em]{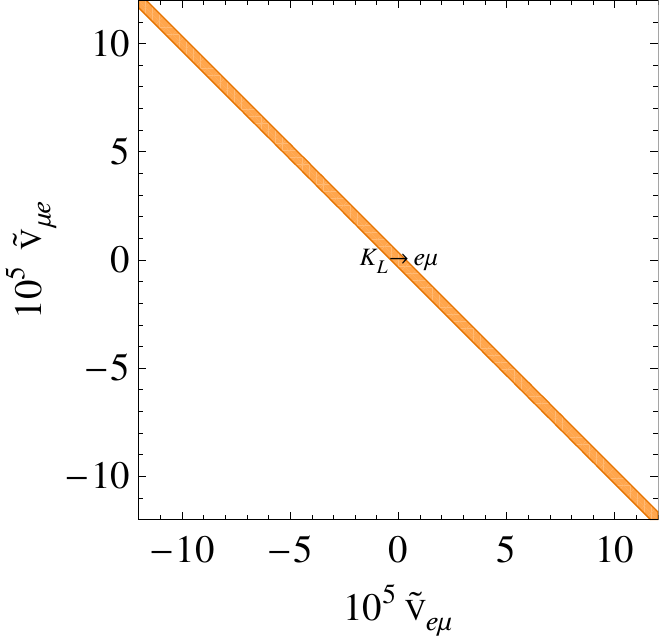}\hspace*{\fill}\vspace{3ex}\\
\hspace*{\fill}%
\includegraphics[width=16em]{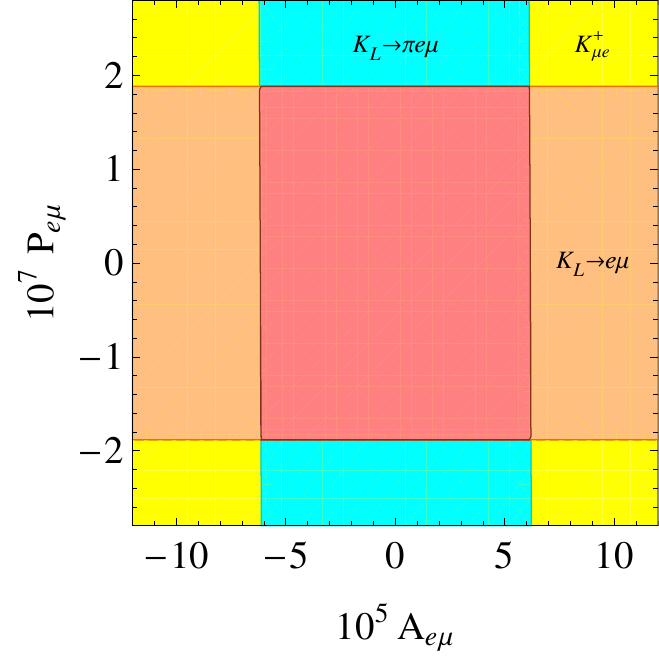} ~ ~ ~ ~
\includegraphics[width=16em]{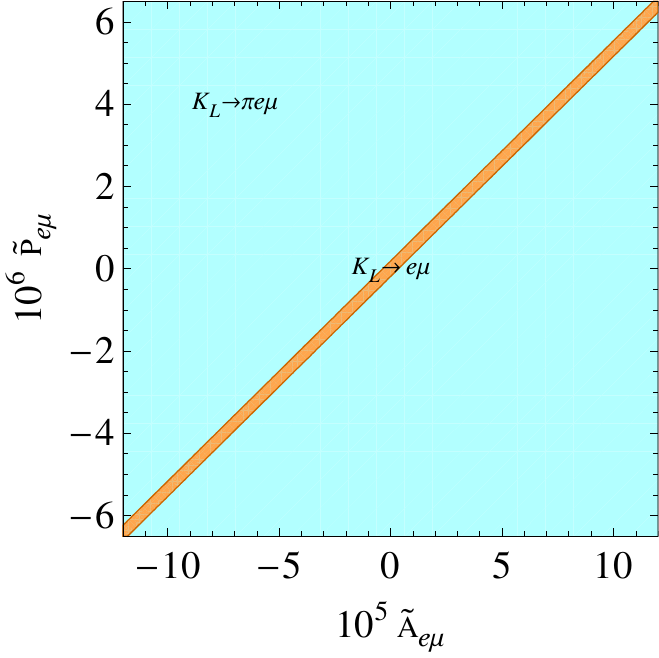}\hspace*{\fill}\vspace{-3pt}
\caption{Regions of $\texttt V_{\mu e}$ versus $\texttt V_{e\mu}$ (top left), $\tilde{\textsc v}_{\mu e}$ versus $\tilde{\textsc v}_{e\mu}$ (top right), $\texttt P_{e\mu}$ versus $\texttt A_{e\mu}$ (bottom left), and $\tilde{\textsc p}_{e\mu}$ versus $\tilde{\textsc a}_{e\mu}$ (bottom right), all taken to be real, for \,$\Lambda_{\textsc{np}}=1$\,TeV,\, allowed by the experimental limits on the branching-fractions of \,$K_L\to\pi^0e^\pm\mu^\mp$, \,$K^+\to\pi^+e^-\mu^+$, \,$K^+\to\pi^+\mu^-e^+$,\, and \,$K_L\to e^\pm\mu^\mp$ $\big($indicated by $K_L\to\pi e\mu$,\, $K_{e\mu}^+$, $K_{\mu e}^+$, and $K_L\to e\mu$, respectively$)$.
In the left (right) plot at the bottom, the bound from \,$K_L\to e^\pm\mu^\mp$ $\big(K_L\to\pi e^\pm\mu^\mp\big)$ is included because they are affected by \,$\tilde{\textsc s}_{e\mu}=-\texttt P_{e\mu}$ $\big(\texttt S_{e\mu}=-\tilde{\textsc p}_{e\mu}\big)$, from eq.\,(\ref{sp}).
In each of the four cases, all the other couplings are set to zero.\label{kaonconstr}\bigskip}
\end{figure}

To illustrate how the different bounds may constrain the couplings, we look at a few examples in which the couplings are real and only two of the independent ones are nonzero at a time.
In figure \ref{kaonconstr}, we display for \,$\Lambda_{\textsc{np}}=1$\,TeV\, the allowed regions of $\texttt V_{e\mu}$ and $\texttt V_{\mu e}$ (top-left plot),  $\tilde{\textsc v}_{e\mu}$ and $\tilde{\textsc v}_{\mu e}$ (top-right plot), $\texttt A_{e\mu}$ and $\texttt P_{e\mu}$ (bottom-left plot), and $\tilde{\textsc a}_{e\mu}$ and $\tilde{\textsc p}_{e\mu}$ (bottom-right plot), subject to the kaon bounds in eq.\,(\ref{Klimits}).
In the bottom-left (-right) plot, the vertical axis implies that $\tilde{\textsc s}_{e\mu}$~$\big(\texttt S_{e\mu}\big)$, which equals \,$-\texttt P_{e\mu}$ $\bigl(-\tilde{\textsc p}_{e\mu}\bigr)$ according to eq.\,(\ref{sp}), is also nonvanishing and consequently influences \,$K_L\to e^\pm\mu^\mp$ $\big(K_L\to\pi^0e^\pm\mu^\mp\big)$, leading to the extra constraint depicted by the orange (light cyan) area on the left (right).

\begin{figure}[t!] \hspace*{\fill}%
\includegraphics[width=16em]{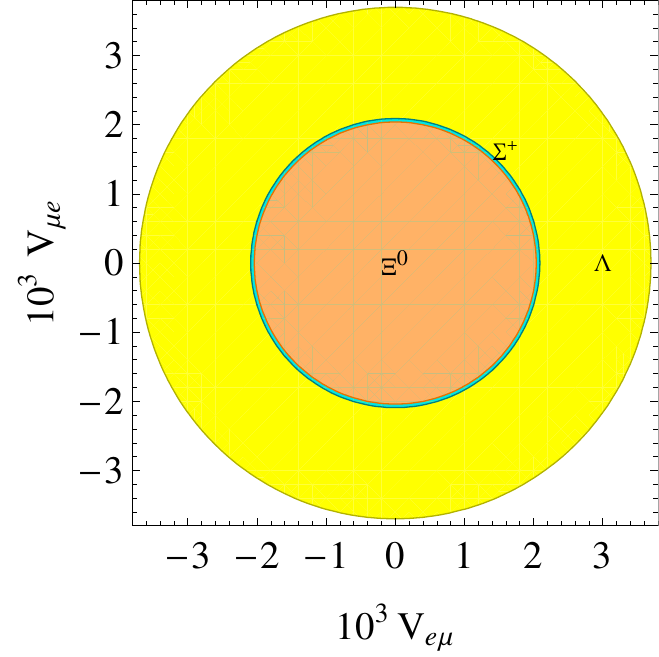} ~ ~ ~ ~
\includegraphics[width=16em]{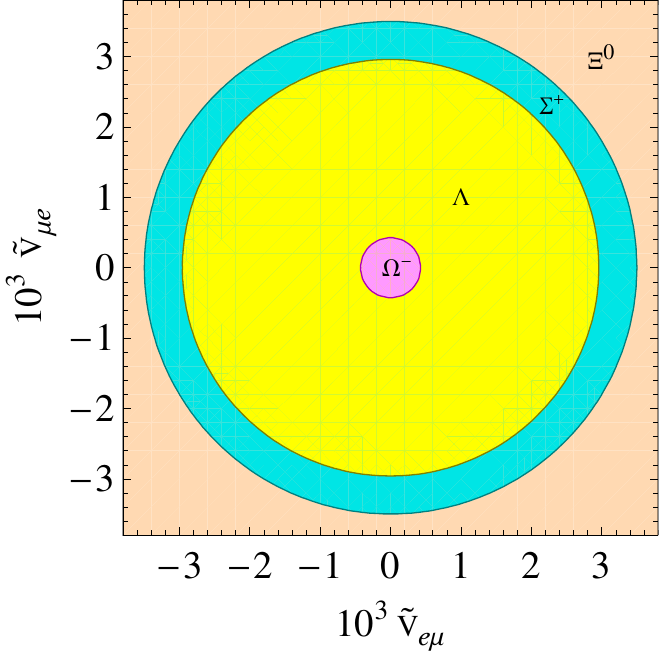}\hspace*{\fill}\vspace{9pt}\\
\hspace*{\fill}%
\includegraphics[width=16em]{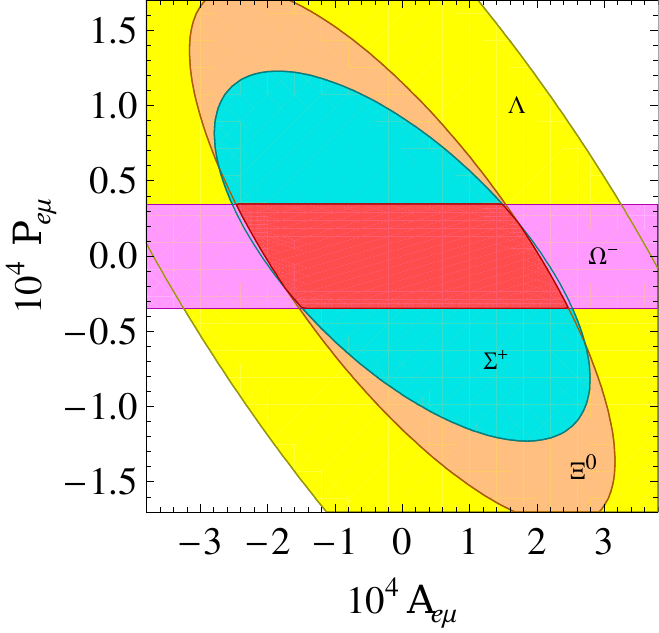} ~ ~ ~ ~
\includegraphics[width=16em]{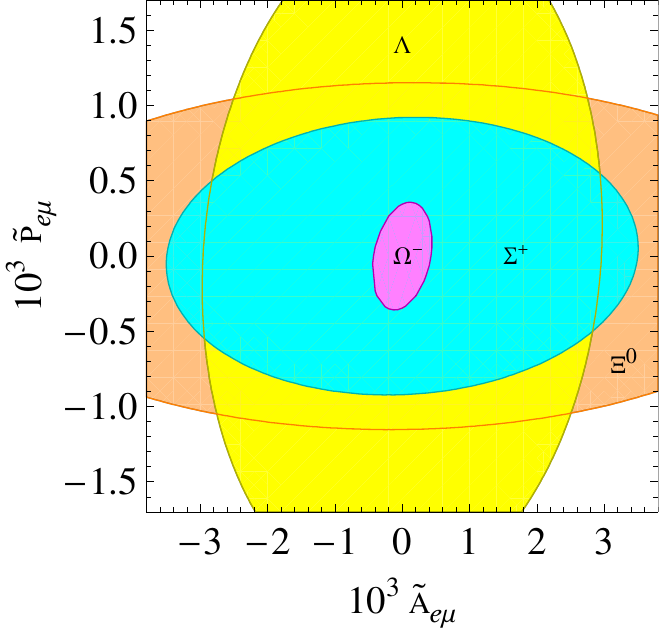}\hspace*{\fill}\vspace{-7pt}
\caption{Allowed regions of $\texttt V_{\mu e}$ versus $\texttt V_{e\mu}$ (top left), $\tilde{\textsc v}_{\mu e}$ versus $\tilde{\textsc v}_{e\mu}$ (top right), $\texttt P_{e\mu}$ versus $\texttt A_{e\mu}$ (bottom left), and $\tilde{\textsc p}_{e\mu}$ and $\tilde{\textsc a}_{e\mu}$ (bottom right), all taken to be real, for \,$\Lambda_{\textsc{np}}=1$\,TeV,\, subject to assumed limits of $10^{-10}$ on the hyperon branching fractions in eqs.\,(\ref{BB2B'em})-(\ref{BO2Xem}), labeled by $\Lambda$, $\Sigma^+$, $\Xi^0$, and $\Omega^-$, respectively.
The bottom plots take into account eq.\,(\ref{sp}).
In each case the other couplings are set to zero.}\label{hypconstr}
\end{figure}

For comparison, given that there are still no direct-search restrictions on hyperon LFV, we entertain the possibility of future experimental limits of $10^{-10}$ on all of the branching fractions in eqs.\,(\ref{BB2B'em})-(\ref{BO2Xem}), inspired by the aforementioned LHCb finding on \,$\Sigma^+\to p\mu^+\mu^-$ \cite{Aaij:2017ddf}.
Under this assumption, we acquire the areas in figure \ref{hypconstr}, which reveals that these constraints are still much weaker than the kaon ones if fine cancelations do not occur among the couplings.
If future hyperon measurements could achieve branching-fraction limits of $10^{-12}$ instead, the allowed regions would be reduced by a factor of 10, from which one can infer that for limits better than $10^{-12}$ the hyperon bounds would start to become comparable to their kaon counterparts.\footnote{The numbers we use to illustrate possible future LHCb bounds are based on the following. Their single event sensitivity (\textsc{ses}) for $\Sigma^+\to p\mu^+\mu^-$ with 3~fb$^{-1}$ is $(2.2\pm 1.2)\times 10^{-9}$ \cite{Aaij:2017ddf}. With expected collection of 50~fb$^{-1}$ in the Phase-I upgrade and assuming that the \textsc{ses} to modes with one muon and one electron is within a factor of five or so, limits of order $10^{-10}$ would be possible. A further collection of 300~fb$^{-1}$ in the Phase-II, combined with expected improvements in trigger efficiency \cite{Junior:2018odx}, leads us to speculate on possible  $10^{-12}$  future limits.}

Before moving on to other transitions without hyperons, here we address how much the NP of interest may influence the determination of input parameters in the SM, particularly the elements of the CKM matrix.
The operators $Q_{2,6,6\prime}^{e\mu,\mu e}$ give rise to interactions involving charged currents and densities, as indicated by the last four rows of table\,\,\ref{tabFR} and partly discussed in subsection \ref{othermodes}, and thus contribute to (semi)leptonic meson decays with a neutrino in the final state that occur already at tree level in the SM but without violating lepton flavor. 
Some of them are among the processes conventionally employed to evaluate the CKM parameters.  
In the presence of $Q_{2,6,6\prime}^{e\mu,\mu e}$, which violate lepton flavor, each of their measured rates would then encompass an increase of order $1/\Lambda_{\textsc{np}}^4$ relative to its SM prediction.  
It follows that the CKM matrix elements extracted from these decays also undergo changes of order $1/\Lambda_{\textsc{np}}^4$, in a way analogous to renormalization of the parameters.
Since the ranges of the associated NP coefficients allowed by the current kaon constraints treated above are very small, as can be deduced from figure~\ref{kaonconstr}, barring major fine-tuning among the coefficients, we conclude that they have negligible effects on the determination of the CKM matrix elements.

\subsection{Other constraints\label{other-constr}}

The branching fractions of other modes that can restrict the NP encoded in eq.\,(\ref{Lnp}) are \cite{Tanabashi:2018oca}
\begin{align}
{\cal B}(\pi^+\to e^+\nu ) & \,=\, (1.230\pm0.004)\times 10^{-4} \,, \nonumber \\
{\cal B}(K^+\to e^+\nu) & \,=\, (1.582\pm0.007)\times 10^{-5} \,, \nonumber \\
{\cal B}(\pi^+\to \mu^+\nu_e) & \,<\, 8.0\times 10^{-3} \,, \nonumber \\
{\cal B}(D_s^+\to e^+\nu) & \,<\, 8.3\times 10^{-5} \,, \nonumber \\
{\cal B}(\pi^0\to e^\pm\mu^\mp) & \,<\, 3.6\times 10^{-10} \,, \nonumber \\{\cal B}(D^0\to e^\pm\mu^\mp) & \,<\, 1.3\times 10^{-8} \,, \nonumber \\
{\cal B}(D^+\to e^+\nu) & \,<\, 8.8\times 10^{-6} \,, \nonumber \\
{\cal B}(D^+\to \pi^+ e^+\mu^-) & \,<\, 2.9\times 10^{-6} \,, \nonumber \\
{\cal B}(D^+\to K^+ e^+\mu^-) & \,<\, 1.2\times 10^{-6} \,,
\end{align}
where the limits are at 90\% CL.
All of these modes supply much weaker constraints than the ones already obtained from the kaon sector in the previous subsection.
An illustrative list is shown in table \ref{t:other} where we compare constraints, at 90\% CL, on the coefficients of (pseudo)scalar operators $Q_{6,6\prime}^{\ell\ell'}$ from only two-body decays, including \,$K_L\to e^\pm\mu^\mp$,\, assuming that \,$c_{6,6\prime}^{\ell\ell'}$\, are equal, real, and the only nonvanishing coefficients.
In the cases where the decays are observed, the limits are obtained from the quoted experimental errors.
For the first four modes in this table, the coupling bounds are computed using eqs.\,\,(\ref{DGM2enu}) and\,\,(\ref{CM}) with the values of decay constants and particle masses from ref.\,\cite{Tanabashi:2018oca} and CKM matrix elements from ref.\,\cite{Charles:2015gya}.
In this particular scenario, with the maximal $\big|c_6^{e\mu}\big|$ from the $K_L$ limit in the last row of the table, the corresponding hyperon branching fractions turn out to be less than \,$2\times10^{-17}$.\,

\begin{table}[b!]  \bigskip
\center{\begin{tabular}{|c|c||c|}
     \hline
\small Process & \small $\begin{array}{c}\rm Upper~limit~at~90\%\,CL ~on~NP~\vspace{-2pt} \\ \rm contribution~to~branching~fraction\end{array}$ & \small Upper bound on \,$\big|c_6^{e\mu}\big|\displaystyle\bigg(\frac{1\rm~TeV}{\Lambda_{\textsc{np}}}\bigg)\raisebox{7pt}{$\!^2$}$
      \\ \hline \hline
 $\pi^+\to e^+\nu$ & $6.6\times 10^{-7} $ & $2.4\times 10^{-3^{\vphantom{o^o}}}$ \\
 $K^+\to e^+\nu$ & $1.2\times 10^{-7} $ & $1.7\times 10^{-4} $ \\
       $D^+\to e^+\nu$ & $8.8\times 10^{-6} $ & $0.037$ \\
        $D_s^+\to e^+\nu$ & $8.3\times 10^{-5} $ & $0.58$ \\ \hline
$K_L\to e^\pm\mu^\mp$ & $4.7\times 10^{-12} $ & $1.9\times 10^{-7^{\vphantom{o^o}}}$ \\ \hline
      \end{tabular}
      \caption{Bounds on coefficients of scalar operators $Q_{6,6\prime}^{\ell\ell'}$ from processes without hyperons under the assumption that $c_{6,6\prime}^{\ell\ell'}$ are equal, real, and the only nonzero coefficients.}
      \label{t:other}}
      \end{table}

A comparison with \,$\mu\to e$\, conversion in nuclei and \,$\pi^0,D^0\to e^\pm\mu^\mp$\, is also instructive.
Since $Q_{6,6\prime}^{\ell\ell'}$ do not affect them, in this instance we suppose that \,$c_{1,5}^{e\mu}=c_{1,5}^{\mu e}$\, and that these coefficients are real and the only ones being nonzero.
Based on eq.\,(\ref{Bm2e}), the existing experimental limits on \,$\mu\to e$\, conversion in various nuclei \cite{Tanabashi:2018oca}, and the corresponding overlap integral and $\omega_{\rm capt}^{\cal N}$ values \cite{Kitano:2002mt}, we expect \,${\cal N}=\rm Au$\, to provide the most consequential constraints.
To evaluate ${\cal B}(\mu^-{\cal N}\to e^-{\cal N})$ for this nucleus, we adopt \,$V_{\rm Au}^{(p)}=0.0974$,\, $V_{\rm Au}^{(n)}=0.146$,\, and \,$\omega_{\rm capt}^{\rm Au}=13.07\times10^6/\rm s$\, from ref.\,\cite{Kitano:2002mt}, and the result is displayed in table \ref{others}.
Therein we also collect the bounds from \,$\pi^0,D^0\to e^\pm\mu^\mp$,\, their rates being given in eq.\,(\ref{D2em}) with \,$c_2^{\ell\ell'}=0$.\,
Evidently, the current data on \,$\mu\to e$\, conversion and \,$K_L\to e^\pm\mu^\mp$\, can yield similarly strong constraints on $c_{1,5}^{\ell\ell'}$.
In this specific case, with the bound from the $K_L$ decay quoted in the last row of the table, we find that the hyperon branching fractions do not exceed \,$4\times10^{-15}$.\,

\begin{table}[h] \bigskip
\center{\begin{tabular}{|c|c||c|}
     \hline
\small Process & \small $\begin{array}{c}\rm Upper~limit~at~90\%\,CL \vspace{-2pt} \\ \rm on~branching~fraction\end{array}$ & \small Upper bound on \,$\sqrt{\big|c_1^{e\mu}\big|\raisebox{1pt}{$^2$}+\big|c_5^{e\mu}\big|\raisebox{1pt}{$^2$}}\displaystyle\bigg(\frac{1\rm~TeV}{\Lambda_{\textsc{np}}}\bigg)\raisebox{7pt}{$\!^2$}$
      \\ \hline \hline
 $\mu^-{\rm Au}\to e^-{\rm Au}$ & $7\times10^{-13}$ & $9.1\times 10^{-6^{\vphantom{o^o}}}$ \\
 $\pi^0\to e^\pm\mu^\mp$ & $3.6\times 10^{-10}$ & $55$ \\
       $D^0\to e^\pm\mu^\mp$ & $1.3\times 10^{-8}$ & $0.050$ \\ \hline
$K_L\to e^\pm\mu^\mp$ & $4.7\times 10^{-12} $ & $5.0\times 10^{-6^{\vphantom{o^o}}}$ \\ \hline
      \end{tabular}
      \caption{Bounds from processes without hyperons on \,$c_{1,5}^{e\mu}=c_{1,5}^{\mu e}$\, if these  coefficients are real and the only ones nonvanishing.}
      \label{others}}
      \end{table}

\section{Concluding remarks\label{conclusions}}

\begin{figure}[h] \bigskip
\center{\includegraphics[width=50mm]{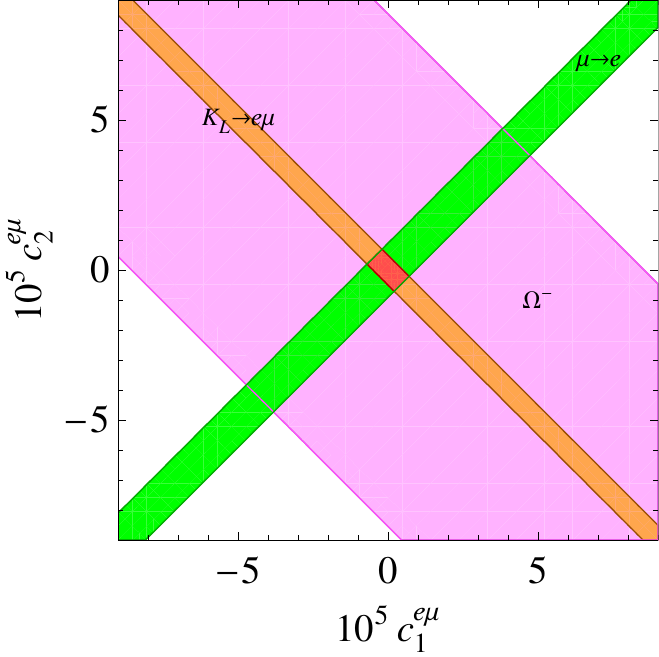} ~
\includegraphics[width=52mm]{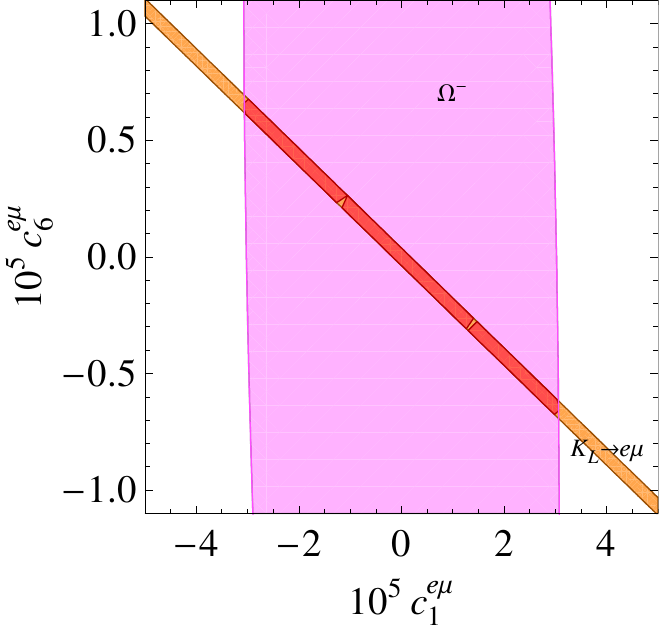} ~
\includegraphics[width=52mm]{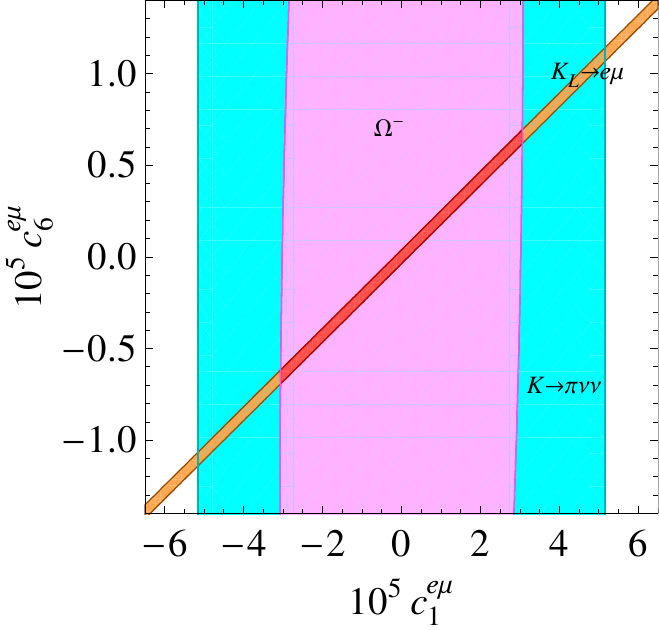}}\vspace{-7pt}
\caption{Comparative constraints on selected LFV couplings in eq.\,(\ref{Lnp}), for \,$\Lambda_{\textsc{np}}=1$\,TeV,\, from current 90\%-CL upper bounds on NP effects in \,$K_L\to \mu^\pm e^\mp$, \,$K^+ \to \pi^+ \nu\bar\nu$,\, and \,$\mu^-\to e^-$\, conversion in gold and a possible future bound of \,${\cal B}\big(\Omega^-\to \Xi^- \mu^\pm e^\mp\big)<10^{-12}$,\, under the general assumption that the couplings are real and \,$c_k^{e\mu}=c_k^{\mu e}$.\,
The specific choices for the nonzero ones are described in the text.}
\label{comparison}
\end{figure}

We have studied charged-lepton-flavor violation in strangeness-changing, $|\Delta S|=1$, transitions, paying special attention to the decays of  hyperons.
We start from the most general effective Lagrangian containing dimension-six operators which are invariant under the SM gauge group and can induce \,$|\Delta S|=1$\, processes with LFV. We illustrate how the operators would appear from the exchange of leptoquarks.
We then explore the contributions of these operators to the hyperon decays as well as their kaon counterparts.
This allows us to contrast the coverage of parameter space that may be achieved in the hyperon sector with what is known from the kaon modes.
In addition, we consider other processes that are affected by the same LFV operators when written in an SU(2)$_L$-gauge-invariant form. Our main results from these comparisons can be summarized as follows.
\begin{itemize}
\item The current experimental exclusion limit on \,$K_L\to\mu^\pm e^\mp$\, places the strongest constraint on LFV operators with a pseudoscalar \,$|\Delta S|=1$\, quark bilinear.
Hyperon decays can only be competitive in this case if an exclusion at the $10^{-16}$ level is reached for the $\Omega^-$ mode.
In the left panel of figure~\ref{comparison1} we illustrate this scenario (the vertical axis) and for the comparison use~\,${\cal B}\big(\Omega^-\to \Xi^-\mu^\pm e^\mp\big)<10^{-12}$.\,
Other hyperon decay modes are even less competitive, as can be deduced from figure~\ref{hypconstr}.
Indirectly, by implication of eq.\,(\ref{sp}), the same can be said of LFV operators with a scalar \,$|\Delta S|=1$\, quark bilinear.

\item
Nevertheless, the left panel of figure~\ref{comparison1} also reveals that in some instances there are combinations of the (pseudo)scalar coefficients which can evade the \,$K_L\to\mu^\pm e^\mp$\, restriction (the horizontal axis in this example) but which can be constrained by the hyperon modes as well as by \,$K\to\pi\mu^\pm e^\mp$.\,
The situation, which is less extreme than that in the preceding scenario, is depicted in the left panel showing that an $\Omega^-$ limit at the $10^{-12}$ level is already starting to be competitive to the currently strictest limit from \,$K^+\to\pi^+e^-\mu^+$.

\item For axial-vector \,$|\Delta S|=1$\, quark bilinears, the situation is also not as extreme and can be seen in the right panel of figure~\ref{comparison1}.
In this case the constraint from \,${\cal B}\big(\Omega^-\to\Xi^-\mu^\pm e^\mp\big)<10^{-12}$\, is only \,$\sim\,$17\, times weaker than the \,$K_L\to\mu^\pm e^\mp$\, one and the hyperons already become competitive at the $10^{-14}$ level.

\item For vector \,$|\Delta S|=1$\, quark bilinears, \,$K_L\to\mu^\pm e^\mp$\, no longer offers a constraint.
Presently the best restrictions on them are from \,$K_L\to\pi^0e^\mp\mu^\pm$\, and \,$K^+\to\pi^+e^-\mu^+$,\, as exhibited in the top-left plot of figure\,\,\ref{kaonconstr}.
Although the $\Omega^-$ mode is insensitive to the vector quark bilinears, the decays of the spin-1/2 hyperons, especially $\Sigma^+$ and $\Xi^0$, can probe them, but branching-fraction limits of order $10^{-13}$ are required to be competitive to the kaon ones, as may be inferred from comparing the top-left plots in figures~\ref{kaonconstr} and~\ref{hypconstr}.

\begin{figure}[t]
\center{\includegraphics[width=17em]{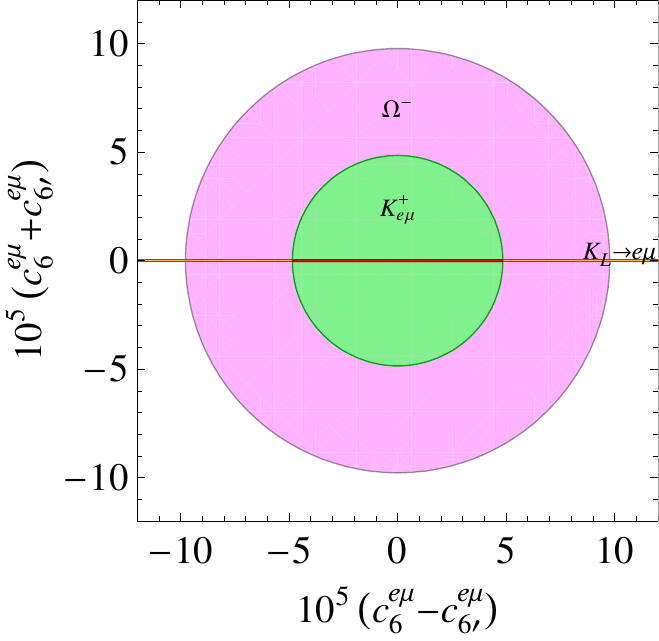} ~ ~ ~ ~
\includegraphics[width=17em]{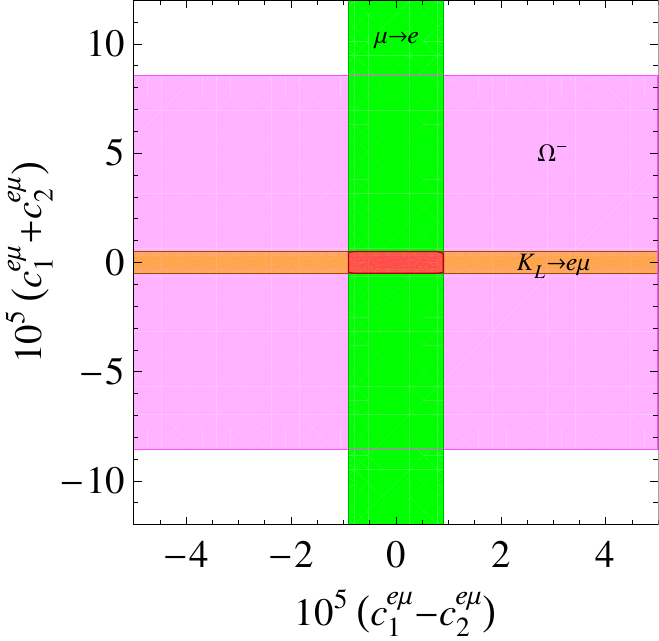}}\vspace{-7pt}
\caption{Comparative constraints on combinations of LFV couplings $c_k^{\ell\ell'}$ from eq.\,(\ref{Lnp}) that produce operators with definite parity, under the assumption that \,$c_k^{e\mu}=c_k^{\mu e}$, they are real, \,$\Lambda_{\textsc{np}}=1$\,TeV, and~\,${\cal B}\big(\Omega^-\to \Xi^- \mu^\pm e^\mp\big)<10^{-12}$.}
\label{comparison1}
\end{figure}

\item The most important constraints from other rare decays correspond to \,$K^+\to \pi^+\nu\bar\nu$\, and \,$\mu^-\to e^-$\, conversion in gold.
Concerning the former, the impact of the couplings is realized via eq.\,(\ref{Wll'}), and so we impose \,$|W_{\ell\ell'}|<2$\, based on the findings of ref.\,\cite{He:2018uey}.
Figures~\ref{comparison} (right panel) and~\ref{comparison1} (right panel) place the limits from these two processes in context.

\item In figure~\ref{comparison}, we illustrate a few selective comparisons of constraints supplied by the different processes.
The specific choices for the nonzero couplings are $c_{1,2}^{\ell\ell'}$ (left panel), \,$c_{1,2}^{\ell\ell'}=c_3^{\ell\ell'}/2$ and $c_6^{\ell\ell'}$ (center panel), and \,$c_{1,2}^{\ell\ell'}=-c_3^{\ell\ell'}/2=c_4^{\ell\ell'}/4$\, and $c_6^{\ell\ell'}$ (right panel).
As this and the previous figures indicate, when all the LFV couplings are present, the different modes complement each other and they all contribute to the overall picture.

\end{itemize}

\section*{Acknowledgements}

This research was supported in part by the MOE Academic Excellence Program (Grant No. 105R891505) and NCTS of ROC.
The work of X.G.H. was supported in part by the MOST of ROC (Grant No. MOST104-2112-M-002-015-MY3 and
106-2112-M-002-003-MY3), in part by the Key Laboratory for Particle Physics, Astrophysics and Cosmology,
Ministry of Education, and Shanghai Key Laboratory for Particle Physics and Cosmology
(Grant No. 15DZ2272100), and in part by the NSFC (Grant Nos. 11575111 and 11735010) of PRC.
G.V. thanks the Physics Department at National Taiwan University for their hospitality and partial support while this work was completed. We thank Jeremy Dalseno  for helpful communications.

\appendix

\section{Feynman Rules}\label{feynrules}

The various four-fermion couplings with (2quark)(2lepton) flavor structures due to $Q_k^{e\mu}$ in eq.\,(\ref{Lnp}) are listed in table~\ref{tabFR}.
Those with the lepton flavors interchanged can be immediately obtained from the corresponding entries in the table by applying the change \,$c_k^{e\mu}\to c_k^{\mu e}$.\,
The Hermitian conjugates of these couplings are additional ones with the quarks interchanged.

\begin{table}[h!] \bigskip
\center{\small
     \begin{tabular}{ | c |c | }
     \hline
$\begin{array}[c]{c}\rm Flavor\vspace{-1ex}\\\rm structure\end{array}$ & Feynman rule
\\ \hline \hline
$(\bar d s) (\bar e\mu)$ &
$\big(c_1^{e\mu}+c_2^{e\mu}\big) L_\eta^{} ${\footnotesize$\otimes$}$ L^\eta
+ c_3^{e\mu} R_\eta^{} ${\footnotesize$\otimes$}$ R^\eta
+ c_4^{e\mu} R_\eta^{} ${\footnotesize$\otimes$}$ L^\eta
+ c_5^{e\mu} L_\eta^{} ${\footnotesize$\otimes$}$ R^\eta
+ c_6^{e\mu} \tilde{\textsl{\texttt L}} \otimes \tilde{\textsl{\texttt R}}
+ c_{6\prime}^{e\mu} \tilde{\textsl{\texttt R}} \otimes \tilde{\textsl{\texttt L}}$
\\  \hline
$(\bar d s) (\bar \nu_e \nu_\mu )$ &
$\big(c_1^{e\mu}-c_2^{e\mu}\big) L_\eta^{} ${\footnotesize$\otimes$}$ L^\eta
      + c_4^{e\mu} R_\eta^{} ${\footnotesize$\otimes$}$ L^\eta$
\\ \hline
$(\bar u u) (\bar e\mu)$ &
$V_{ud}^{}V_{us\,}^* \big[ \big(c_1^{e\mu} - c_2^{e\mu}\big) L_\eta^{} $$\otimes$$ L^\eta
+ c_5^{e\mu} L_\eta^{} $$\otimes$$ R^\eta \big]$
\\ \hline
$(\bar u c) (\bar e \mu)$ &
$V_{ud}^{}V_{cs\,}^* \big[ \big(c_1^{e\mu}-c_2^{e\mu}\big) L_\eta^{} ${\footnotesize$\otimes$}$ L^\eta + c_5^{e\mu} L_\eta^{} ${\footnotesize$\otimes$}$ R^\eta \big]$
\\  \hline
$(\bar u u) (\bar \nu_e \nu_\mu)$ &
$V_{ud}^{}V_{us\,}^* \big(c_1^{e\mu}+c_2^{e\mu}\big) L_\eta^{} ${\footnotesize$\otimes$}$ L^\eta$
\\  \hline
$(\bar u c) (\bar \nu_e \nu_\mu)$ &
$V_{ud}^{}V_{cs\,}^* \big(c_1^{e\mu}+c_2^{e\mu}\big) L_\eta^{} ${\footnotesize$\otimes$}$ L^\eta$
\\  \hline
$(\bar d u) (\bar\nu_e \mu)$ & $V_{us\,}^* \big( 2 c_2^{e\mu} L_\eta^{} ${\footnotesize$\otimes$}$ L^\eta + c_6^{e\mu}\, \tilde{\textsl{\texttt L}} \otimes \tilde{\textsl{\texttt R}} \big)$
\\  \hline
$(\bar d c) (\bar\nu_e\mu)$
      & $V_{cs\,}^* \big( 2c_2^{e\mu} L_\eta^{} ${\footnotesize$\otimes$}$ L^\eta
      + c_6^{e\mu}\, \tilde{\textsl{\texttt L}} \otimes \tilde{\textsl{\texttt R}} \big)$
\\  \hline
 $ (\bar u s) (\bar e \nu_\mu )$ & $V_{ud}^{}\, \big( 2 c_2^{e\mu} L_\eta^{} ${\footnotesize$\otimes$}$ L^\eta + c_{6\prime}^{e\mu}\, \tilde{\textsl{\texttt R}} \otimes \tilde{\textsl{\texttt L}} \big)$
\\  \hline
 $ (\bar c s) (\bar e \nu_\mu )$ & $V_{cd}^{}\, \big( 2 c_2^{e\mu} L_\eta^{} ${\footnotesize$\otimes$}$ L^\eta + c_{6\prime}^{e\mu}\, \tilde{\textsl{\texttt R}} \otimes \tilde{\textsl{\texttt L}} \big)$
      \\ \hline
      \end{tabular}
\caption{Feynman rules arising from $Q_k^{e\mu}$ in eq.\,(\ref{Lnp}).
In the second column, each entry is to be multiplied by $i/\Lambda_{\textsc{np}}^2$ and completed with the Dirac spinors of the fermions in the first column, we have defined \,$L_\eta^{}=\gamma_\eta^{}P_L^{}$,\, $R_\eta^{}=\gamma_\eta^{}P_R^{}$,\, $\tilde{\textsl{\texttt L}}=P_L^{}$,\, and \,$\tilde{\textsl{\texttt R}}=P_R^{}$,\, and the element $V_{\textsl{\texttt U}_i\textsl{\texttt D}_j}$ corresponds to $({\cal V}_{\textsc{ckm}})_{ij}^{}$ in eq.\,(\ref{ql}).
The neutrinos being nearly massless and unobserved in decays, we display their weak eigenstates \,$\nu_e^{}=({\cal U}_{\textsc{pmns}})_{1j}^{}\nu_{jL}'$\, and \,$\nu_\mu^{}=({\cal U}_{\textsc{pmns}})_{2j}^{}\nu_{jL}'$\, in the first column.
}
      \label{tabFR}
}
      \end{table}

\section{Correspondences between quark and hadron transitions\label{correspondences}}

From the chiral Lagrangian which is at lowest order in the derivative and $s$-quark-mass
($m_s$) expansions and describes the strong interactions among the lightest octet baryons
and mesons and decuplet baryons~\cite{Gasser:1983yg,Bijnens:1985kj,Jenkins:1991es},
one can extract correspondences between quark densities or currents and hadronic
transitions \cite{He:2005we}.
From the results of ref.\,\cite{He:2005we} pertaining to the \,$|\Delta S|=1$\, processes
under discussion, one can infer \cite{Tandean:2019tkm}
\begin{align} \label{corresp}
\bar d\gamma_\eta^{}s & \;\Leftrightarrow\; -\sqrt{\frac{3}{2}}~\overline{n}\gamma_\eta^{}\Lambda
- \overline{p}\gamma_\eta^{}\Sigma^+
+ \sqrt{\frac{3}{2}}~ \overline{\Lambda}\gamma_\eta^{}\Xi^0
- \frac{1}{\sqrt2}\, \overline{\Sigma^0} \gamma_\eta^{}\Xi^0
+ \overline{\Sigma\bar{\hphantom{o}}}\gamma_\eta^{}\Xi^-
\nonumber \\ & ~~~ ~~ +\,
i \big( \pi^+\, \partial_\eta K^- - K^-\, \partial_\eta \pi^+ \big)
- \frac{i}{\sqrt2} \big( \pi^0\, \partial_\eta \overline{K}{}^0
- \overline{K}{}^0\, \partial_\eta \pi^0 \big)
+\, \cdots \,, & \hspace{5ex}
\end{align}
\begin{align}
\bar d s & \;\Leftrightarrow\;
\sqrt{\frac{3}{2}}~\frac{m_\Lambda^{}-m_N^{}}{\hat m-m_s^{}}\, \overline{n} \Lambda
+ \frac{m_\Sigma^{}-m_N^{}}{\hat m-m_s^{}}\, \overline{p}\,\Sigma^+
+ \sqrt{\frac{3}{2}}~ \frac{m_\Xi^{}-m_\Lambda^{}}{m_s^{}-\hat m}\, \overline{\Lambda}\,\Xi^0
\nonumber \\ & ~~~ ~~ +\,
\frac{m_\Xi^{}-m_\Sigma^{}}{\hat m-m_s^{}} \Bigg( \frac{\overline{\Sigma^0}\, \Xi^0}{\sqrt2}
- \overline{\Sigma\bar{\hphantom{o}}}\,\Xi^- \Bigg)
+\, B_0^{} \Bigg( \pi^+K^- - \frac{\pi^0 \overline{K}{}^0}{\sqrt2} \Bigg) +\, \cdots \,, &
\end{align}
\begin{align} \label{dgg5s}
\bar d\gamma_\eta^{}\gamma_5^{}s & \;\Leftrightarrow\;
\frac{-D-3F}{\sqrt6}~ \overline{n}\gamma_\eta^{}\gamma_5^{}\Lambda
+ (D-F)\, \overline{p}\gamma_\eta^{}\gamma_5^{}\Sigma^+
- \frac{D-3F}{\sqrt6}~ \overline{\Lambda}\gamma_\eta^{}\gamma_5^{}\Xi^0   \hspace{7em}
\nonumber \\ & ~~~ ~~ -\,
\frac{D+F}{\sqrt2}~ \overline{\Sigma^0}\gamma_\eta^{}\gamma_5^{}\Xi^0 + (D+F)\,
\overline{\Sigma\bar{\hphantom{o}}}\gamma_\eta^{}\gamma_5^{}\Xi^-
\,+\, {\cal C}\, \overline{\Xi\bar{\hphantom{o}}}\, \Omega_\eta^-
\,+\, \sqrt2\, f\, \partial_\eta^{} \overline{K}{}^0
\,+\, \cdots \,,
\end{align}
\begin{align} \label{dg5s}
\bar d\gamma_5^{}s & \;\Leftrightarrow\; i\sqrt2\, B_0^{}\, f \overline{K}{}^0 \,+\, \cdots \,,  \hspace{23em}
\end{align}
where $m_{N,\Sigma,\Xi}^{}$ are isospin-averaged masses of the nucleons, $\Sigma^{\pm,0}$, and $\Xi^{0,-}$, respectively, $\hat m$ is the average mass of the $u$ and $d$ quarks, \,$B_0=m_K^2/(\hat m+m_s)$,\, with $m_K$ here being the average mass of $K^0$ and $K^-$, the free parameters $D$, $F$, and $\cal C$ occur in the leading-order chiral Lagrangian and can be fixed from baryon decay data, \,$f=f_K^{}/\sqrt2$,\, and the ellipses represent terms irrelevant to our analysis.

At the same order in the chiral expansion, the baryonic matrix elements of
\,$\bar d\big(\gamma^\eta,1\big)\gamma_5^{}s$\, also receive contributions from kaon-pole diagrams involving
\,$\langle0|\bar d\big(\gamma^\eta,1\big)\gamma_5^{}s|\overline{K}{}^0\rangle$\, from eqs.\,\,(\ref{dgg5s}) and\,\,(\ref{dg5s}) and vertices from the lowest-order strong chiral Lagrangian ${\cal L}_{\rm s}$.
In the latter, the pertinent terms are given by \cite{Tandean:2019tkm}
\begin{align}
{\cal L}_{\rm s}^{} \,\supset &~ \Bigg[
\frac{-D-3F}{\sqrt6}~ \overline{n} \gamma_\eta^{}\gamma_5^{}\Lambda
+ (D-F)\, \overline{p} \gamma_\eta^{}\gamma_5^{}\Sigma^+
- \frac{D-3F}{\sqrt6}~ \overline{\Lambda} \gamma_\eta^{}\gamma_5^{}\Xi^0
\nonumber \\ & ~~ -\,
\frac{D+F}{\sqrt2}~ \overline{\Sigma^0} \gamma_\eta^{}\gamma_5^{}\Xi^0
+ (D+F)\, \overline{\Sigma\bar{\hphantom{o}}} \gamma_\eta^{}\gamma_5^{}\Xi^-
+\, {\cal C}\, \overline{\Xi\bar{\hphantom{o}}}\, \Omega_\eta^-
\Bigg] \frac{\partial^\eta K^0}{\sqrt2\, f} \,.
\end{align}
From this and the preceding paragraphs, we arrive at the matrix elements in eqs.\,\,(\ref{<B'B>}), (\ref{<X-O->}), (\ref{<vacK>}), and (\ref{<pi-K->}) in the limit that \,$f_{+,0}^{}=1$.\,

Numerically, we adopt \,$D=0.81$\, and \,$F=0.46$\, determined from fitting to the data on hyperon semileptonic decays and \,${\cal C}=1.7$\, from the measurements of strong decays of the decuplet spin-3/2 baryons into an octet spin-1/2 baryon and a pion \cite{Tanabashi:2018oca}.\footnote{With this $\cal C$ value and the differential rate in eq.\,(\ref{G'O2Xem}) suitably modified for \,$s\to u\bar\nu e^-$\, in the SM, we can predict \,${\cal B} (\Omega^-\to\Xi^0\bar\nu e^-)_{\textsc{sm}}^{}\simeq0.60\%$\, in agreement with its measurement \cite{Tanabashi:2018oca}.}
Furthermore, we use the measured hadron masses from ref.\,\cite{Tanabashi:2018oca} and, for light meson and hyperon decays, the light-quark mass values \,$\hat m=(m_u^{}+m_d^{})/2=4.4$ MeV\, and \,$m_s^{}=120$ MeV\, at a renormalization scale of~1~GeV. 
These quark masses have been rescaled from their values at a renormalization scale of~2~GeV available from ref.\,\cite{Tanabashi:2018oca}, which are also employed in subsection \ref{other-constr} for treating the charmed meson decays.

\section{Additional kaon decay formulas\label{Kformulas}}

With the matrix elements in eq.\,(\ref{<vacK>}), for the \,${\cal K}\to\ell^-\ell^{\prime+}$
amplitude in eq.\,(\ref{MK2ll'}) we obtain
\begin{align}
S_{\overline K{}^0 e\mu}^{} & = -S_{K^0\mu e}^* = \frac{f_K^{}}{\Lambda_{\textsc{np}}^2} \big(
\tilde{\textsc v}_{e\mu\,} m_\mu^{} + B_{0\,}^{} \tilde{\textsc s}_{e\mu}^{} \big) \,, &
P_{\overline K{}^0 e\mu}^{} & =  P_{K^0\mu e}^* = \frac{-f_K^{}}{\Lambda_{\textsc{np}}^2} \big(
\tilde{\textsc a}_{e\mu\,} m_\mu^{} - B_{0\,}^{} \tilde{\textsc p}_{e\mu}^{} \big) \,,
\nonumber \\
S_{\overline K{}^0\mu e}^{} & = -S_{K^0 e\mu}^* = \frac{-f_K^{}}{\Lambda_{\textsc{np}}^2} \big(
\tilde{\textsc v}_{\mu e\,}^{} m_\mu^{} - B_{0\,}^{} \tilde{\textsc s}_{\mu e}^{} \big) \,, &
P_{\overline K{}^0\mu e}^{} & =  P_{K^0 e\mu}^* = \frac{-f_K^{}}{\Lambda_{\textsc{np}}^2} \big(
\tilde{\textsc a}_{\mu e\,}^{} m_\mu^{} - B_{0\,}^{} \tilde{\textsc p}_{\mu e}^{} \big) \,.
\end{align}
Employing the approximate relations \,$\sqrt2\, K_{L,S}= K^0\pm\overline K{}^0$,\, we then find
\begin{align}
S_{K\!_L^{}e\mu}^{} & \,=\, -S_{K\!_L^{}\mu e}^* \,=\, \frac{f_K^{}}{\sqrt2\, \Lambda_{\textsc{np}}^2}
\Big[ \big(\tilde{\textsc v}_{e\mu}^{}+\tilde{\textsc v}{}_{\mu e}^*\big) m_\mu^{}
+ B_0^{} \big(\tilde{\textsc s}_{e\mu}^{}-\tilde{\textsc s}{}_{\mu e}^*\big) \Big] \,, &
\nonumber \\
P_{K\!_L^{}e\mu}^{} & \,=\, P_{K\!_L^{}\mu e}^* \,=\, \frac{f_K^{}}{\sqrt2\, \Lambda_{\textsc{np}}^2}
\Big[ \bigl(-\tilde{\textsc a}_{e\mu}^{}-\tilde{\textsc a}{}_{\mu e}^*\bigr) m_\mu^{}
+ B_0^{}\big(\tilde{\textsc p}_{e\mu}^{}+\tilde{\textsc p}{}_{\mu e}^*\big) \Big] \,,
\\
S_{K\!_S^{}e\mu}^{} & \,=\, S_{K\!_S^{}\mu e}^* \,=\, \frac{f_K^{}}{\sqrt2\, \Lambda_{\textsc{np}}^2}
\Big[ \bigl(-\tilde{\textsc v}_{e\mu}^{}+\tilde{\textsc v}{}_{\mu e}^*\bigr) m_\mu^{}
- B_0^{} \big(\tilde{\textsc s}_{e\mu}^{} + \tilde{\textsc s}{}_{\mu e}^*\big) \Big] \,, &
\nonumber \\
P_{K\!_S^{}e\mu}^{} & \,=\, -P_{K\!_S^{}\mu e}^* \,=\, \frac{f_K^{}}{\sqrt2\, \Lambda_{\textsc{np}}^2}
\Big[ \big(\tilde{\textsc a}_{e\mu}^{}-\tilde{\textsc a}{}_{\mu e}^*\big) m_\mu^{}
- B_0^{}\big(\tilde{\textsc p}_{e\mu}^{}-\tilde{\textsc p}{}_{\mu e}^*\big) \Big] \,,
\end{align}
which go into eq.\,(\ref{GK2em}).

With the kaon-to-pion matrix elements from subsection \ref{kaons}, for \,$K^-\to\pi^-\mu^\mp e^\pm$\, and their antiparticle counterparts the $S$ and $P$ terms in eq.\,(\ref{MK2pll'}) are
\begin{align}
\Lambda_{\textsc{np}\,}^2 S_{K^-\pi^-\mu e}^{} & = \Big[ (f_-^{}-f_+^{}) m_\mu^{}
+ 2 f_+^{}\, \slashed p_{\!K} \Big] \texttt V_{\mu e}^{} + B_0^{} f_0^{} \texttt S_{\mu e}^{} \,, &
P_{K^-\pi^-\mu e}^{} & = S_{K^-\pi^-\mu e}
\big|^{\texttt V_{\mu e}\to\texttt A_{\mu e}}_{\texttt S_{\mu e}\to\texttt P_{\mu e}} \,,
\nonumber \\
\Lambda_{\textsc{np}\,}^2 S_{K^-\pi^-e\mu}^{} & = \Big[ (f_+^{}-f_-^{}) m_\mu^{}
+ 2 f_+^{}\, \slashed p_{\!K} \Big] \texttt V_{e\mu}^{} + B_0^{} f_0^{} \texttt S_{e\mu}^{} \,, &
P_{K^-\pi^-e\mu}^{} & = S_{K^-\pi^-\mu e}
\big|^{\texttt V_{\mu e}\to\texttt A_{e\mu}}_{\texttt S_{\mu e}\to\texttt P_{e\mu}} \,,
\\
\Lambda_{\textsc{np}\,}^2 S_{K^+\pi^+\mu e}^{} & = \Bigl[ (f_+^{}-f_-^{}) m_\mu^{}
- 2 f_+^{}\, \slashed p_{\!K} \Bigr] \texttt V_{e\mu}^* + B_0^{}f_0^{} \texttt S_{e\mu}^* \,, &
P_{K^+\pi^+\mu e}^{} & = S_{K^+\pi^+\mu e}
\big|^{\texttt V_{e\mu}\to\texttt A_{e\mu}}_{\texttt S_{e\mu}\to-\texttt P_{e\mu}} \,,
\nonumber \\
\Lambda_{\textsc{np}\,}^2 S_{K^+\pi^+e\mu}^{} & = \Bigl[ (f_-^{}-f_+^{}) m_\mu^{}
- 2 f_+^{}\, \slashed p_{\!K} \Bigr] \texttt V_{\mu e}^* + B_0^{} f_0^{} \texttt S_{\mu e}^* \,, &
P_{K^+\pi^+e\mu}^{} & = S_{K^+\pi^+\mu e}
\big|^{\texttt V_{e\mu}\to\texttt A_{\mu e}}_{\texttt S_{e\mu}\to-\texttt P_{\mu e}} \,,
\end{align}
where \,$f_-^{}=\big(f_0^{}-f_+^{}\big)\big(m_K^2-m_\pi^2\big)/\hat s$.\,
Moreover, given that
\,${\cal M}_{K^-\to\pi^-e^\pm\mu^\mp}=-\sqrt2\, {\cal M}_{\bar K^0\to\pi^0e^\pm\mu^\mp}$ and
\,${\cal M}_{K^+\to\pi^+e^\pm\mu^\mp}=-\sqrt2\, {\cal M}_{K^0\to\pi^0e^\pm\mu^\mp}$,\,
for the analogous decays of $K_L$ and $K_S$
\begin{align}
\Lambda_{\textsc{np}\,}^2 S_{K_{\!L,S\,}^{}\pi^0\mu e}^{} & \,=\, \Bigl[ \tfrac{1}{2}
\big(f_-^{}-f_+^{}\big)m_\mu^{} + f_+^{}\, \slashed p_{\!K} \Big] \texttt V_\mp^{}
\,-\, \tfrac{1}{2} B_0^{} f_0^{}\, \texttt S_\pm^{} \,,
\nonumber \\
\Lambda_{\textsc{np}\,}^2 P_{K_{\!L,S\,}^{}\pi^0\mu e}^{} & \,=\, \Bigl[ \tfrac{1}{2}
\big(f_-^{}-f_+^{}\big)m_\mu^{} + f_+^{}\, \slashed p_{\!K} \Big] \texttt A_\mp^{}
\,+\, \tfrac{1}{2} B_0^{} f_0^{}\, \texttt P_\mp^{} \,,
\\
\Lambda_{\textsc{np}\,}^2 S_{K_{\!L,S\,}^{}\pi^0e\mu}^{} & \,=\, \mp \Big[ \tfrac{1}{2}
\big(f_+^{}-f_-^{}\big) m_\mu^{} + f_+^{}\,\slashed p_{\!K} \Big] \texttt V_\mp^*
\,\mp\, \tfrac{1}{2} B_0^{} f_0^{}\, \texttt S_\pm^* \,,
\nonumber \\
\Lambda_{\textsc{np}\,}^2 P_{\!K_{L,S\,}^{}\pi^0e\mu}^{} & \,=\, \mp \Big[ \tfrac{1}{2}
\big(f_-^{}-f_+^{}\big) m_\mu^{} + f_+^{}\,\slashed p_{\!K} \Big] \texttt A_\mp^*
\,\mp\, \tfrac{1}{2} B_0^{} f_0^{}\, \texttt P_\mp^* \,,
\end{align}
where
\begin{align}
\texttt V_\pm^{} & \,=\, \texttt V_{e\mu}^* \pm \texttt V_{\mu e}^{} \,, &
\texttt S_\pm^{} & \,=\, \texttt S_{e\mu}^* \pm \texttt S_{\mu e}^{} \,, &
\texttt A_\pm^{} & \,=\, \texttt A_{e\mu}^* \pm \texttt A_{\mu e}^{} \,, &
\texttt P_\pm^{} & \,=\, \texttt P_{e\mu}^* \pm \texttt P_{\mu e}^{} \,.
\end{align}
The presence of $\slashed p{}_{\!K}^{}$ in $P_{{\cal K}\pi\ell\ell'}^{}$ implies that $P_{{\cal K}\pi\ell\ell'}^{}\gamma_5^{}$ in eq.\,(\ref{MK2pll'}) is not the same as $\gamma_5^{}P_{{\cal K}\pi\ell\ell'}^{}$.
From these $S_{{\cal K}\pi\ell\ell'}$ and $P_{{\cal K}\pi\ell\ell'}$ formulas follow the differential decay rates\footnote{In this study we ignore the possibility that the coupling parameters could have both strong and weak phases.
Otherwise, the decay rates of a pair of $CP$-conjugate modes would generally be different, leading to $CP$-violating rate asymmetries.}
\begin{align} \label{G'K-}
\Gamma_{K^-\to\pi^-\mu^-e^+}' \,=\, \Gamma & _{K^+\to\pi^+e^-\mu^+}'
\nonumber \\ =\,
\frac{\beta^4 \lambda_{K^+\pi^+\,}^{1/2} f_0^2}{64\pi^3 m_{K^+}^3 \Lambda_{\textsc{np}}^4} &
\Bigg[ \Bigg( \frac{3-\beta^2}{6f_0^2}\, \lambda_{K^+\pi^+\,}^{} f_+^2 + \frac{\Delta_{K^+\pi^+}^4
\, m_\mu^2}{2\hat s} \Bigg) \big(|\texttt V_{\mu e}|^2+|\texttt A_{\mu e}|^2\big)
\nonumber \\ & \,+\,
\Delta_{K^+\pi^+}^2\, B_0^{}\, m_\mu^{}\, {\rm Re} \big( \texttt A_{\mu e}^* \texttt P_{\mu e}^{}
+ \texttt V_{\mu e}^* \texttt S_{\mu e}^{} \big) + \frac{B_0^{2\,} \hat s}{2}
\big(|\texttt S_{\mu e}|^2+|\texttt P_{\mu e}|^2\big) \Bigg] \,,
\end{align}
\begin{align}
\Gamma_{K^-\to\pi^-e^-\mu^+}' \,=\, \Gamma & _{K^+\to\pi^+\mu^-e^+}'
\nonumber \\ =\,
\frac{\beta^4 \lambda_{K^+\pi^+\,}^{1/2} f_0^2}{64\pi^3 m_{K^+}^3 \Lambda_{\textsc{np}}^4} &
\Bigg[ \Bigg( \frac{3-\beta^2}{6 f_0^2}\, \lambda_{K^+\pi^+\,}^{} f_+^2 + \frac{\Delta_{K^+\pi^+}^4
\, m_\mu^2}{2\hat s} \Bigg) \big(|\texttt V_{e\mu}|^2+|\texttt A_{e\mu}|^2\big)
\nonumber \\ & \,+\,
\Delta_{K^+\pi^+}^2\, B_0^{}\, m_\mu^{}\, {\rm Re} \big( \texttt A_{e\mu}^* \texttt P_{e\mu}^{}
- \texttt V_{e\mu}^* \texttt S_{e\mu}^{} \big) + \frac{B_0^{2\,} \hat s}{2}
\big(|\texttt S_{e\mu}|^2+|\texttt P_{e\mu}|^2\big) \Bigg] \,,
\end{align}
\begin{align}
\Gamma_{K_L^{}\to\pi^0\mu^-e^+}' \,=\, \Gamma_{K_L^{}\to}' & _{\pi^0e^-\mu^+}^{}
\nonumber \\ ~=\,
\frac{\beta^4 \lambda_{K^0\pi^0\,}^{1/2}f_0^2}{256\pi^3m_{K^0}^3\Lambda_{\textsc{np}}^4} &
\Bigg[ \Bigg( \frac{3-\beta^2}{6 f_0^2}\, \lambda_{K^0\pi^0\,}^{} f_+^2 + \frac{\Delta_{K^0\pi^0}^4
\, m_\mu^2}{2\hat s} \Bigg) \big( |\texttt V_-|^2 + |\texttt A_-|^2 \big)
\nonumber \\ & \,+\,
\Delta_{K^0\pi^0}^2\, B_0^{}\, m_\mu^{}\, {\rm Re} \big( \texttt A_-^* \texttt P_-^{}
- \texttt V_-^* \texttt S_+^{} \big) + \frac{B_0^{2\,} \hat s}{2}
\big( |\texttt S_+|^2 + |\texttt P_-|^2 \big) \Bigg] \,,
\end{align}
\begin{align} \label{G'KS}
\Gamma_{K_S^{}\to\pi^0\mu^-e^+}' \,=\, \Gamma_{K_S^{}\to}' & _{\pi^0e^-\mu^+}^{}
\nonumber \\ ~=\,
\frac{\beta^4 \lambda_{K^0\pi^0\,}^{1/2}f_0^2}{256\pi^3m_{K^0}^3\Lambda_{\textsc{np}}^4} &
\Bigg[ \Bigg( \frac{3-\beta^2}{6 f_0^2}\, \lambda_{K^0\pi^0\,}^{} f_+^2 + \frac{\Delta_{K^0\pi^0}^4
\, m_\mu^2}{2\hat s} \Bigg) \big( |\texttt V_+|^2 + |\texttt A_+|^2 \big)
\nonumber \\ & \,+\,
\Delta_{K^0\pi^0}^2\, B_0^{}\, m_\mu^{}\, {\rm Re} \big( \texttt A_+^* \texttt P_+
- \texttt V_+^* \texttt S_-^{} \big) + \frac{B_0^{2\,} \hat s}{2}
\big( |\texttt S_-|^2 + |\texttt P_+|^2 \big) \Bigg] \,,
\end{align}
where \,$\Delta_{XY}^2=m_X^2-m_Y^2$.\,

\bibliography{biblio}

\end{document}